\documentclass[pre,onecolumn,showpacs]{revtex4-1}

\usepackage{amsmath,amssymb,amsbsy}
\usepackage{graphicx}
\usepackage[colorlinks]{hyperref}

\usepackage{todonotes}

\newcommand{\BB}{\mathcal{B}} 
\newcommand{\be}{\mathbf{e}} 
\newcommand{\bH}{\mathbf{H}} 
\newcommand{\He}{H_{\rm e}} 
\newcommand{\bHe}{\bH_{\rm e}} 
\newcommand{\bM}{\mathbf{M}} 
\newcommand{\Msat}{M_{\rm sat}}
\newcommand{\TT}{T_\omega}
\newcommand{\bxi}{\boldsymbol{\xi}} 
\newcommand{\xie}{\xi_\mathrm{e}} 
\newcommand{\bOmega}{\boldsymbol{\Omega}} 
\newcommand{\tauB}{\tau_{\rm B}}
\newcommand{\kT}{k_{\rm B}T}
\newcommand{\dd}{\mathrm{d}}

\begin{document}

\title{Analytic solution for the nonlinear response of magnetic nanoparticles to large-amplitude oscillatory fields}

\author{Patrick Ilg}

\affiliation{School of Mathematical, Physical, and Computational Sciences, University of Reading, Reading, RG6 6AX, United Kingdom}

\date{\today}

\begin{abstract}
Nonlinear responses of physical systems to strong perturbations are notoriously difficult to tackle analytically.
Here, we present analytic results for the nonlinear response of magnetic nanoparticles to large-amplitude oscillatory magnetic fields based on a particular model for the magnetization dynamics.
A number of characteristic features of the in-phase and out-of-phase higher-harmonic response are found and analyzed.
In particular we find that the magnitude of higher harmonic contributions $R_n$ depends on the field amplitude and frequency only via a single scaling variable that combines the two quantities.
The decrease of $|R_n|$ with increasing order $n$ of harmonics is a key quantity monitored 
in biomedical applications such as magnetic particle spectroscopy and magnetic particle imaging. 
Except for the first few harmonics, we find that this decrease is exponential with a rate that depends on the scaling variable only. 
For not too high frequencies and not too large amplitudes, we find that these exact results for one particular model of magnetization dynamics hold approximately also for other, more frequently used models. 
Our results therefore offer not only deeper insight into strongly nonlinear responses of magnetic nanoparticles, especially for higher harmonics that are very difficult to determine numerically, but also suggest analyzing data in terms of a scaling variable. 
\end{abstract}

\maketitle

\section{Introduction}

Learning about a system from its response to perturbations is a very successful approach in physics that has also proven to be extremely fruitful for various applications. 
For weak perturbations, the powerful machinery of linear response theory shows that a single function encapsulates  the system's response to arbitrary perturbations provided they are weak enough  \cite{Gaspard_irreversible2022}. 
The dynamic magnetic susceptibility is an example of such a response function that determines the magnetization resulting from weak, time-varying magnetic fields.
Therefore, the dynamic susceptibility is an extremely important quantity to characterize magnetic materials and is deservedly studied in great detail \cite{Odenbach_LNP763,Ferguson:2013gi}.
However, the dynamic susceptibility is applicable only for weak magnetic fields, while many applications probe responses beyond the linear regime.

Magnetic nanoparticles (MNP) and ferrofluids are particularly well-suited model systems to study nonlinear responses because their very large magnetic moments lead to superparamagnetic behavior where even moderate field strengths can exceed the linear regime \cite{Odenbach_LNP763}.
The strong response of MNPs and ferrofluids to external magnetic fields makes these systems interesting as adaptive materials and enables various engineering applications \cite{kole_engineering_2021}.
The response of MNPs to time-varying magnetic fields is also exploited in a number of emerging biomedical applications \cite{rivera-rodriguez_emerging_2021}.
As a step toward accounting for nonlinear responses in a systematic manner, the
very recently proposed Medium Amplitude Field Susceptometry (MAFS) \cite{ilg_medium_2024} introduces a general third-order susceptibility to extend the linear regime up to moderate field strengths.
%

While the medium-amplitude range studied in MAFS can still be characterized by a general third-order susceptibility governing systems' responses to arbitrary time-dependent fields, this is no longer the case for magnetic fields with large amplitudes where the highly non-linear response depends on details of the strong perturbation.
Such strongly nonlinear responses are notoriously difficult to calculate analytically and challenging to comprehensively cover in simulations.
Consequently, very few general results are available for systems subject to strong perturbations. 
Noteworthy exceptions are fluctuation theorems that apply to rather general nonequilibrium situations \cite{Gaspard_irreversible2022}, 
which have interesting consequences for MNPs such as negative heat production in oscillating magnetic fields \cite{ilg_stochastic_2024}. 
In the vast majority of cases, however, we  rely on numerical studies of particular models and certain choices of strong perturbations and selected parameter values.

Here, we consider the nonlinear response of MNPs subject to large-amplitude oscillatory magnetic fields.
This response is of particular importance e.g.\ in Magnetic Particle Spectroscopy (MPS) and Magnetic Particle Imaging (MPI), where 
the magnitude of higher-harmonics of the response encodes important information on the local environment of the MNPs  \cite{wu_magnetic_2020,yang_applications_2022,harvell-smith_magnetic_2022,vogel_critical_2022}.
Detailed experiments have shown that interaction effects between MNPs can have profound effects on the MPS signal  \cite{moor_particle_2022}.
Nevertheless, for a better understanding of the nonlinear response, we here focus on noninteracting MNPs.
To date, only few numerical studies of the nonlinear response of MNPs to large-amplitude oscillatory magnetic fields are available  \cite{felderhof_nonlinear_2003,Yoshida2009,kuznetsov_nonlinear_2022,draack_multiparametric_2019,enpuku_harmonic_2023,zverev_computer_2021}.
Most of these studies are based on the model of Brownian rotations of noninteracting rigid dipoles originally proposed by Martsenyuk \emph{et al.}\ (MRS) \cite{MRS74}, while Zverev \emph{et al.}\ \cite{zverev_computer_2021} simulate the Langevin dynamics of interacting rigid dipoles. 
Even focusing on  the noninteracting case, 
so far (mostly due to numerical limitations) only the first 5 \cite{Yoshida2009,kuznetsov_nonlinear_2022}, 7 \cite{enpuku_harmonic_2023} and 11  \cite{draack_multiparametric_2019} harmonics were studied, 
while experimentally determined higher harmonics extend up to $n_{\rm max} \approx 30\ldots 80$. 
Analytical approaches provide a very useful alternative that is free of such limitations and therefore provide a more complete picture.
Unfortunately, the adiabatic approximation is the only analytical approach to this problem available so far, which is restricted to   sufficiently low excitation frequencies that the system always remains in equilibrium at the instantaneous field value.
Adopting this approximation, the behavior of the in-phase response with increasing field strength has been obtained for all harmonics \cite{barrera_magnetization_2022}.
The same approach was used earlier e.g.\ to test suitability of MNPs for MPI applications \cite{biederer_magnetization_2009}.

While these studies have certainly helped to improve our understanding of the nonlinear response of noninteracting rigid dipoles, neglecting out-of-phase or dissipative contributions as well as the restriction to very small oscillation frequencies are severe limitations of the theory.
Here, we propose a different approach and obtain the exact solution for the nonlinear response including all higher harmonics of the in-phase and out-of-phase contributions valid for all frequencies and up to intermediate field amplitudes.
It is important to note that these results are obtained for a particular model of the nonlinear magnetization dynamics proposed by Shliomis in 2001 (Sh01) \cite{Shlio01}.
There is a long-standing discussion in the literature about the appropriate form of the magnetization equation for ferrofluids (see e.g.\ the chapter by M.\ Liu in Ref.\ \cite{Odenbach_LNP763} and references in Ref.\ \cite{Shlio01}). 
For our purposes, we interpret Sh01 as a more tractable version of the MRS model and provide  
comparisons of the analytical results for the Sh01 model to numerical results obtained for MRS model and its effective field approximation (EFA). 
%

\section{Model equations and exact solution}

Although a plethora of rival magnetization equations have been proposed to describe the dynamics of MNPs and ferrofluids, no consensus on the appropriate form has been achieved so far, except maybe for the ultra-dilute, rigid-dipole case (see e.g.\ \cite{Odenbach_LNP763,Shlio01} and references therein). 
Comparisons of several of these magnetization equations  to experiments on rotating ferrofluids and on magnetoviscosity have not been fully conclusive, due at least in part to uncertainties about fluid characterization and appropriate parametrization of the models  \cite{EmbsLuecke_rigidrot,weng_magnetoviscosity_2013}. 

\subsection{Sh01 model}

A relatively simple magnetization equation that is valid also far from equilibrium and was found to perform well in comparison to  experiments \cite{weng_magnetoviscosity_2013} was proposed by Shliomis in 2001 \cite{Shlio01}. 
Within the Sh01 model, the nonequilibrium magnetization is given by 
\begin{equation} \label{M_t}
\bM(t) = \Msat L(\xie(t)) \hat{\bxi}_{\rm e}(t), 
\end{equation}
where $\Msat$ denotes the saturation magnetization and $L(x)=\coth(x)-1/x$ the Langevin function. 
Equation \eqref{M_t} is formally identical to the equilibrium relation with the magnetic field replaced by the effective field $\bHe$, 
$\bxi_{\rm e}=m\bHe/\kT$ denotes the dimensionless effective field, $\xie=|\bxi_{\rm e}|$ its norm and 
$\hat{\bxi}_{\rm e}=\bxi_{\rm e}/\xie$ the corresponding unit vector. 
The magnetic moment of an individual MNP and the thermal energy are denoted by $m$ and $\kT$, respectively. 
The effective field in the Sh01 model obeys the phenomenological equation 
\begin{equation} \label{Sh01_full} 
\frac{\dd\bHe}{\dd t} -\bOmega\times\bHe = - \frac{1}{\tau}(\bHe - \bH) - \frac{1}{6\eta\phi} \bHe \times  ( \bM \times\bH ) ,
\end{equation}
with $\bOmega$ the flow vorticity, $\tau$ the relaxation time, and $\eta$ and $\phi$ the fluid viscosity and volume fraction of magnetic material, respectively. 

Here, we consider oscillatory magnetic fields with given frequency $\omega$ and amplitude $H$, 
\begin{equation} \label{H_osci}
\bH(t) = H \cos(\omega t) \hat{\bH} . 
\end{equation} 
The unit vector $\hat{\bH}=\bH/H$ denotes the direction of the applied field. 
Under quiescent conditions, $\bOmega=0$, the magnetization response is parallel to the applied field. 
Therefore, Eq.\ \eqref{Sh01_full} reduces to a simple relaxation equation for the effective field that can be solved exactly to give 
\begin{equation} \label{He_t}
\bHe(t) = \frac{1}{\tau} \int_{-\infty}^t e^{-(t-t')/\tau}\bH(t')\dd t' .
\end{equation}
The solution \eqref{He_t} for oscillating fields \eqref{H_osci} can be written as 
\begin{equation} \label{He_sol}
\He(t) = \frac{H}{1+(\omega\tau)^2}\Big( \cos(\omega t) + \omega\tau\sin(\omega t) \Big) . 
\end{equation} 
With the solution \eqref{He_sol} at hand, the magnetization is given by Eq.\ \eqref{M_t} with 
$\xie(t)=m\He(t)/\kT$ and $\hat{\bxi}_{\rm e}(t)=\hat{\bH}$. 
We note that this solution holds for arbitrary frequencies $\omega$ and field amplitudes $H$. 
Also note that the solution \eqref{He_t} corresponds to the situation where the field has been applied indefinitely. 
If instead the field is switched on at a finite time in the past, $t_0<t$, initial transients proportional to $e^{-(t-t_0)/\tau}$ will appear in Eq.\ \eqref{He_sol}.  

\subsection{Fourier series}

For field amplitudes $H$ exceeding the linear response regime, the resulting magnetization $M(t)$ is in general not linearly related to $H$. 
However, $M(t)$ is a periodic function in time with period $\TT=2\pi/\omega$. 
Therefore, Fourier's theorem allows us to always express the resulting magnetization as 
\begin{equation} \label{M_Fourier}
M(t) = M_0 + \sum_{n=1}^\infty\Big( A_n\cos(n\omega t) + B_n\sin(n\omega t) \Big) , 
\end{equation}
where due to symmetry only odd harmonics $n$ contribute to the sum \cite{felderhof_nonlinear_2003}. 
Here and in the following, we concentrate on the magnetization component parallel to the field direction, $M(t)=\bM(t)\cdot\hat{\bH}$.
Note that periodicity of $M(t)$ and therefore Eq.\ \eqref{M_Fourier} is valid in practice for times large enough such that initial  transients have sufficiently decayed to be negligible. 
The coefficients $A_n, B_n$ in Eq.\ \eqref{M_Fourier} can be obtained from the signal $M(t)$ by integration, 
\begin{equation} \label{An_integral}
A_n = \frac{2}{\TT}\int_0^{\TT} M(t)\cos(n\omega t)\dd t, \quad 
B_n = \frac{2}{\TT}\int_0^{\TT} M(t)\sin(n\omega t)\dd t . 
\end{equation}
For oscillating fields \eqref{H_osci}, the time-averaged magnetization $M_0$ vanishes and the magnetization response \eqref{M_Fourier} is determined by the field- and frequency-dependent coefficients $A_n, B_n$ for all odd $n$. 

In MPS and MPI experiments, typically the magnitude $|R_n|=\sqrt{A_n^2+B_n^2}$ and phase angle $\tan\phi_n=B_n/A_n$ of the higher harmonics are measured as a function of order $n$ for selected field strengths $H$ and frequencies $\omega$ \cite{wu_magnetic_2020,barrera_magnetization_2022}. 

\subsection{Exact solution for higher harmonics} \label{exact.sec}

To obtain better insight into the highly nonlinear response, we aim to express the analytical solution \eqref{M_t}  with Eq.\ \eqref{He_sol} for the Sh01 model in the form \eqref{M_Fourier}. 
In this way, explicit expressions for the Fourier coefficients $A_n, B_n$ can be obtained. 

We start with the series expansion of the Langevin function, 
\begin{equation} \label{L_series}
L(x) =  \sum_{n=1}^\infty \frac{2^{2n}\BB_{2n}x^{2n-1}}{(2n)!} , 
\end{equation}
where $\BB_n$ are Bernoulli numbers \cite{gradsteyn}. Convergence of the sum in Eq.\ \eqref{L_series} is guaranteed only for $|x|<\pi$.
To overcome this limitation, different representations of the Langevin function have been considered in Ref.\ \cite{barrera_magnetization_2022}. 
However, due to the complicated expressions arising in these cases, the analysis was limited to the adiabatic approximation, leading to $B_n=0$ in Eq.\ \eqref{M_Fourier} for all $n$. 

Here, we use Eq.\ \eqref{L_series} and accept the limitation $|h|<\pi$, where $h=mH/\kT$ denotes the dimensionless field amplitude.  Inserting Eq.\ \eqref{L_series} into \eqref{M_t} and substituting $\xie(t)$ from Eq.\ \eqref{He_t}, the magnetization is obtained as a power series expression in $\cos(\omega t)$ and $\sin(\omega t)$. 
To make further progress, we use the identity 
\begin{equation} \label{coshochn}
 \Big( \cos(\omega t) + \omega\tau \sin(\omega t) \Big)^{2n-1} 
 = \sum_{k=0}^{n-1} C_{nk}\cos((2n-2k-1)\omega t) + \sum_{k=0}^{n-1} S_{nk}\sin((2n-2k-1)\omega t) , 
\end{equation}
where 
\begin{align}
 C_{nk} & = D_{nk} \sum_{\nu=0; {\rm even}}^{2n-2k-1} \binom{2n-2k-1}{\nu}(-1)^{\nu/2}(\omega\tau)^\nu \\
 S_{nk} & =  D_{nk} \sum_{\nu=1; {\rm odd}}^{2n-2k-1} \binom{2n-2k-1}{\nu}(-1)^{(\nu-1)/2}(\omega\tau)^\nu
\end{align}
and 
$D_{nk}=2^{-2n+2}\binom{2n-1}{k}(1+(\omega\tau)^2)^k$. 
With the help of relation \eqref{coshochn} the magnetization is given by 
\begin{equation}
 M (t)  =  M_{\rm sat}  \sum_{n=1}^\infty \frac{2^{2n}\BB_{2n}h^{2n-1}}{(2n)![1+(\omega\tau)^2]^{2n-1}} \sum_{k=0}^{n-1} \Big[  C_{nk}\cos((2n-2k-1)\omega t) + S_{nk}\sin((2n-2k-1)\omega t) \Big] .
\end{equation} 
Collecting all terms with the same frequency, the magnetization is finally expressed in the form \eqref{M_Fourier} with $M_0=0$ and Fourier coefficients 
\begin{align}
  A_n & = \Msat R_n(h,\omega\tau) \cos(n\tan^{-1}(\omega\tau)) , \label{An_Sh01}\\
  B_n & = \Msat R_n(h,\omega\tau) \sin(n\tan^{-1}(\omega\tau)) , \label{Bn_Sh01}
\end{align} 
for odd $n$, where the functions 
\begin{equation}
R_n(h,\omega\tau) =   \sum_{k=0}^\infty  
 \frac{8(-1)^{(n+2k+3)/2}}{(2\pi)^{n+2k+1}}\binom{n+2k}{k}\zeta(n+2k+1)  
 \left(\frac{h}{\sqrt{1+(\omega\tau)^2}} \right)^{n+2k}  
 \label{Rn_zeta}
\end{equation}
denote their dimensionless magnitude, $|R_n|=\sqrt{A_n^2 + B_n^2}/\Msat$.
In Eq.\ \eqref{Rn_zeta}, the Riemann zeta function $\zeta(z)$ 
has been introduced which is related to the Bernoulli numbers by 
$\BB_{2n} = \frac{(-1)^{n+1}2(2n)!}{(2\pi)^{2n}}\zeta(2n)$ \cite{gradsteyn}. 

Equations \eqref{An_Sh01} -- \eqref{Rn_zeta} constitute one of the main results of this paper. 
They present the exact expressions for all higher harmonic responses to oscillating magnetic fields within the Sh01 model for any frequency and any field amplitude up to $|h|<\pi$. 
A number of comments are in order. 
(i) As expected, only odd harmonics $n$ contribute to the sum in Eq.\ \eqref{M_Fourier}. 
(ii) In-phase contributions $A_n$ are even and out-of-phase contributions $B_n$ are odd functions of frequency. 
(iii) In the low-frequency limit, $\omega\to 0$, we find $A_n\to\Msat R_n(h,0)$ and $B_n\to 0$, as expected from the adiabatic approximation. 
(iv) In the opposite limit of very high frequencies, $\omega\tau\gg 1$, we find 
$A_n \approx R_n\cos(n\pi/2)$, 
$B_n \approx  R_n\sin(n\pi/2)$, 
where $R_n$ approaches zero very fast for increasing frequency. 
(v) Fourier coefficients $A_n$ and $B_n$ of the $n$th harmonic can be expressed as a power series in the field amplitude  starting with $h^n$. 
(vi) Successive contributions $A_n, A_{n+2}$ and $B_n, B_{n+2}$ have alternating signs at low frequencies since $R_n\sim(-1)^{(n+2k+3)/2}$.
(vii) The coefficients $A_n$ and $B_n$ change sign $(n-1)/2$ times as a function of frequency. 
(viii) The dependence of the magnitude $R_n$ on the field amplitude $h$ and frequency $\omega$ in Eq.\ \eqref{Rn_zeta} can be described by a single scaling variable $h/\sqrt{1+(\omega\tau)^2}$.
(ix) Although the infinite sum in Eq.\ \eqref{Rn_zeta} cannot be expressed in closed form, a useful approximate expression for higher harmonics can be derived by approximating $\zeta(n+2k+3)\approx 1$. For $n= 7$ the error is less than $1\%$ and decreases quickly with increasing $n$.
Accepting this approximation we find
\begin{equation} \label{Rn_scal}
 R_n(h,\omega\tau) \approx (-1)^{(n+3)/2}\frac{4}{\pi}\,\frac{z^n}{\sqrt{1+z^2}[1+\sqrt{1+z^2}]^n} \quad\text{for}\ 
 n\gtrsim 7,
\end{equation}
with the scaling variable
\begin{equation}
 z = \frac{h}{\pi\sqrt{1+(\omega\tau)^2}} .
\end{equation}
(x) The loss angle defined by $\tan\phi_n=B_n/A_n$ is found from Eqs.\ \eqref{An_Sh01} and \eqref{Bn_Sh01} to be given by
\begin{equation} \label{phi_n}
\tan\phi_n = \tan(n \tan^{-1}(\omega\tau)) \quad\Rightarrow 
\phi_n = k\pi + n\tan^{-1}(\omega\tau), \quad k\in\mathbb{N}. 
\end{equation}
Thus, $\phi_n$ is determined only up to multiples of $\pi$. With the convention $-\pi\leq\phi_n\leq\pi$, $\phi_n$ is monotonically increasing with frequency $\omega$ and harmonic number $n$ 
for $n\tan^{-1}(\omega\tau)<\pi$, then jumping down. 
It is remarkable that $\phi_n$ is independent of the magnetic field amplitude $h$ in this model. 

To make contact with previous results, 
we consider the exact relations \eqref{An_Sh01} -- \eqref{Rn_zeta}  in the limit of weak field amplitudes, $h\ll 1$. 
In this case, we find from Eq.\ \eqref{Rn_zeta} 
$R_1=h/[3\sqrt{1+(\omega\tau)^2}]+{\cal O}(h^3)$ using $\zeta(2)=\pi^2/6$ and therefore the familiar Debye behavior 
$A_1=\Msat h/[3(1+(\omega\tau)^2)]$ and 
$B_1=\Msat h\omega\tau/[3(1+(\omega\tau)^2)]$ is recovered. 
Similarly, but keeping higher-order terms in the expansion, the leading third-order response is found from Eqs.\ \eqref{An_Sh01} -- \eqref{Rn_zeta} as 
\begin{equation}
A_3 = M_{\rm sat} \frac{-[1-3(\omega\tau)^2]h^3}{180[1+(\omega\tau)^2]^3} + {\cal O}(h^5), \quad
B_3 = M_{\rm sat} \frac{-[3(\omega\tau)-(\omega\tau)^3]h^3}{180[1+(\omega\tau)^2]^3} + {\cal O}(h^5) .
\end{equation}
These expressions agree with Eqs.\ (B.9), (B.10)  in Ref.\ \cite{ilg_medium_2024}, which were obtained from the general third-order susceptibility when applied to the  Sh01 model.


\section{Illustrations and Comparisons} \label{illust.sec}

\subsection{Results for the Sh01 model}
\begin{figure}[htbp]
\begin{center}
\includegraphics[width=0.45\textwidth]{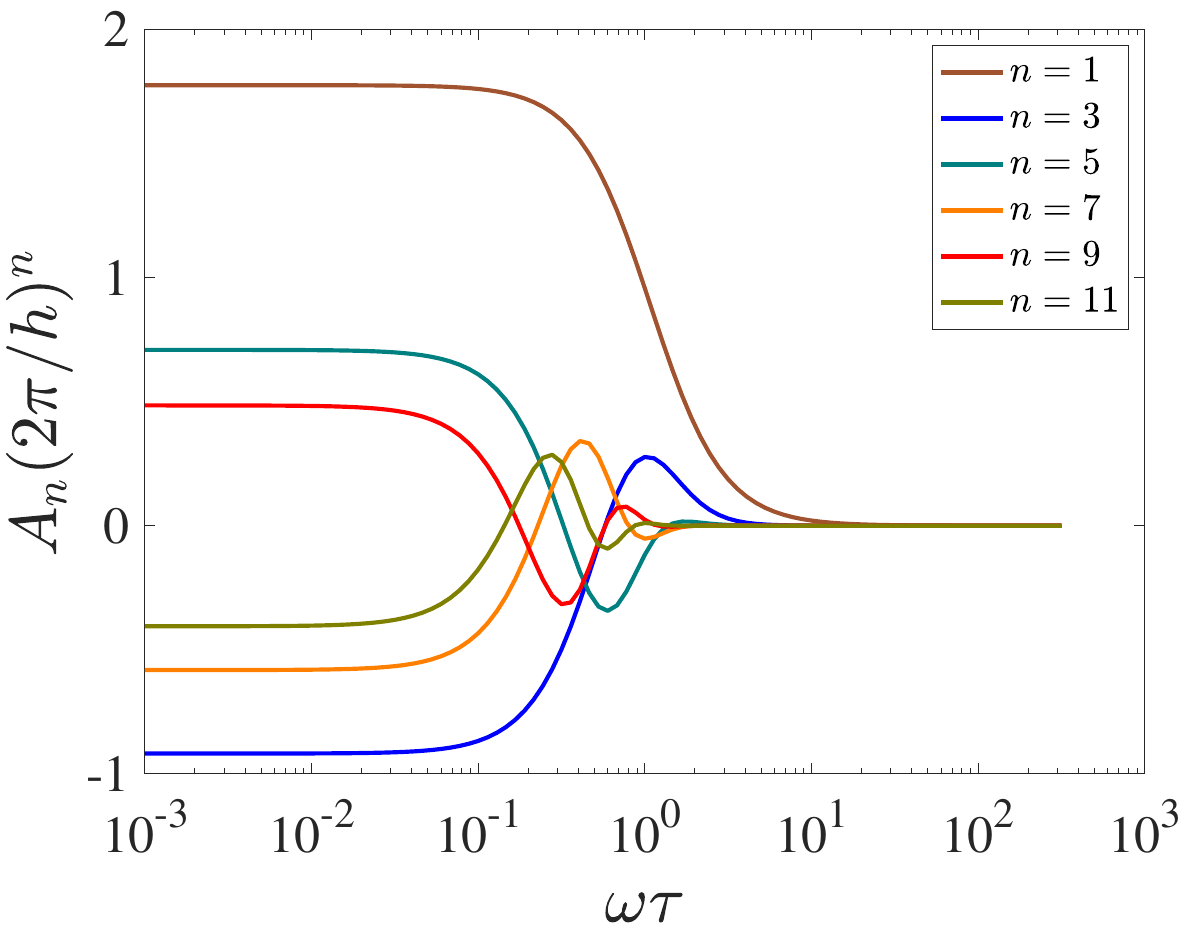}
\includegraphics[width=0.45\textwidth]{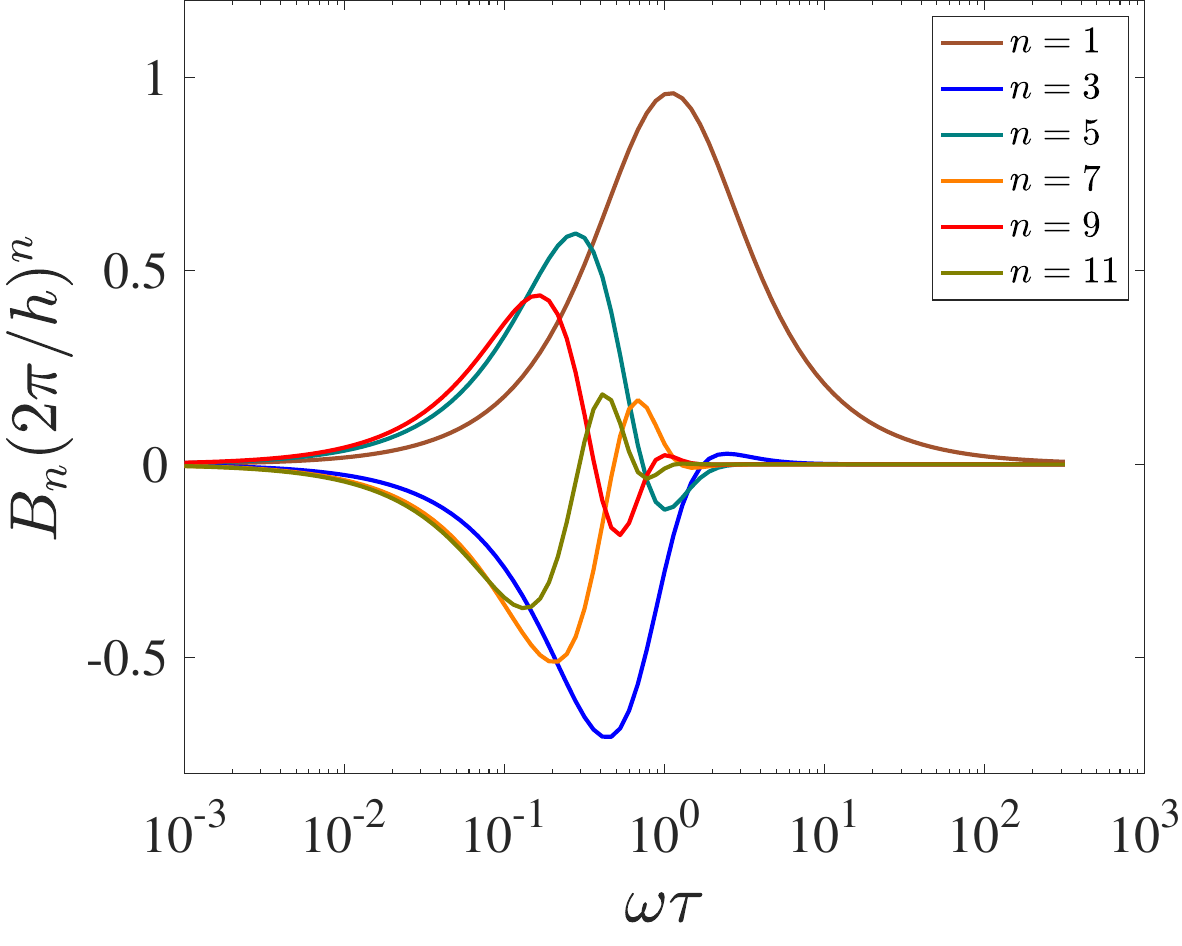}
\caption{Fourier coefficients $A_n$ (left) and $B_n$ (right) of the in-phase and out-of-phase contributions in Eq.\ \eqref{M_Fourier} versus dimensionless frequency $\omega\tau$ for the Sh01 model for the first six odd harmonics $n=1,\ldots,11$. 
The dimensionless magnetic field was chosen as $h=2$. 
For better visibility, $A_n$ and $B_n$ are each scaled with $(2\pi/h)^n/\Msat$ and semi-logarithmic axes are used.}
\label{AnBn_w.fig}
\end{center}
\end{figure}

To further unpack aspects of the exact solution, here we illustrate this result in a number of ways. 
To begin,  Fig.\ \ref{AnBn_w.fig} shows the first 11 Fourier coefficients $A_n$ and $B_n$ from Eq.\  \eqref{An_Sh01} and \eqref{Bn_Sh01} as a function of frequency for a fixed amplitude ($h=2$).   
As mentioned previously, only odd $n$ appear in the expansion. 
Furthermore,  $A_n$ and $B_n$ can be written as power series in the field amplitude starting with $h^n$. 
For better visibility, we therefore scale in Fig.\ \ref{AnBn_w.fig} the Fourier coefficients by a factor $(2\pi/h)^n$ to at least partially account for the strong decrease in their magnitude with increasing order of harmonics $n$. 

From Fig.\ \ref{AnBn_w.fig}, we find that both, $A_n$ and $B_n$ decay rapidly  for high frequencies with the high-frequency limit setting in already for $\omega\tau\gtrsim 10$ for higher harmonics ($n\geq 3$). 
We also observe that $A_n$ approaches its low-frequency limit already for $\omega\tau\lesssim 10^{-2}$. 
On the other hand, the approach of $B_n$ to zero in the low-frequency limit is slower and extends to significantly smaller frequencies.  
This finding indicates that corrections to the adiabatic approximation can become noticeable already at rather low frequencies. 
We also see the alternating signs of $A_n$ and $B_n$ at low frequencies (remark (vi)). 
The number of zero crossings follows remark (vii), even though this becomes difficult to observe for large $n$ on the scale shown in  Fig.\ \ref{AnBn_w.fig}. 

\begin{figure}[htbp]
\begin{center}
\includegraphics[width=0.32\textwidth]{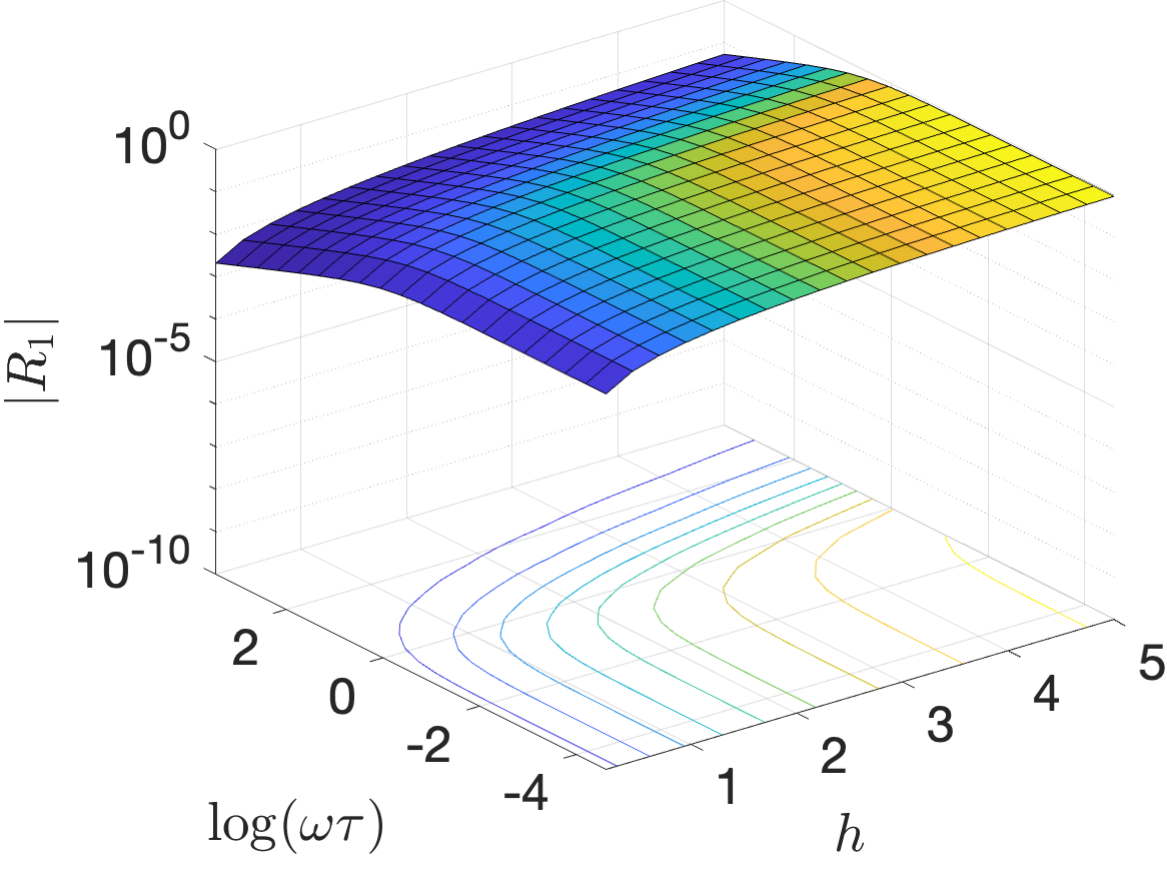}
\includegraphics[width=0.32\textwidth]{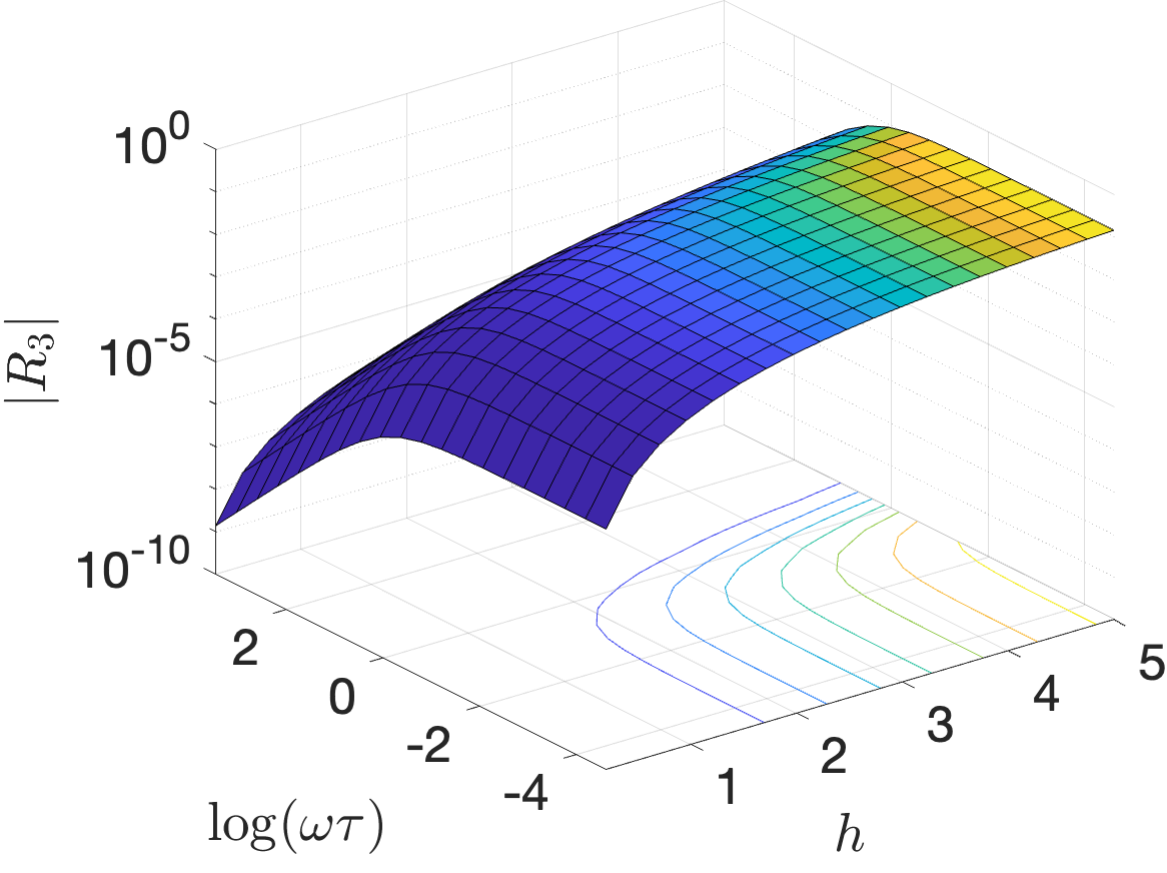}
\includegraphics[width=0.32\textwidth]{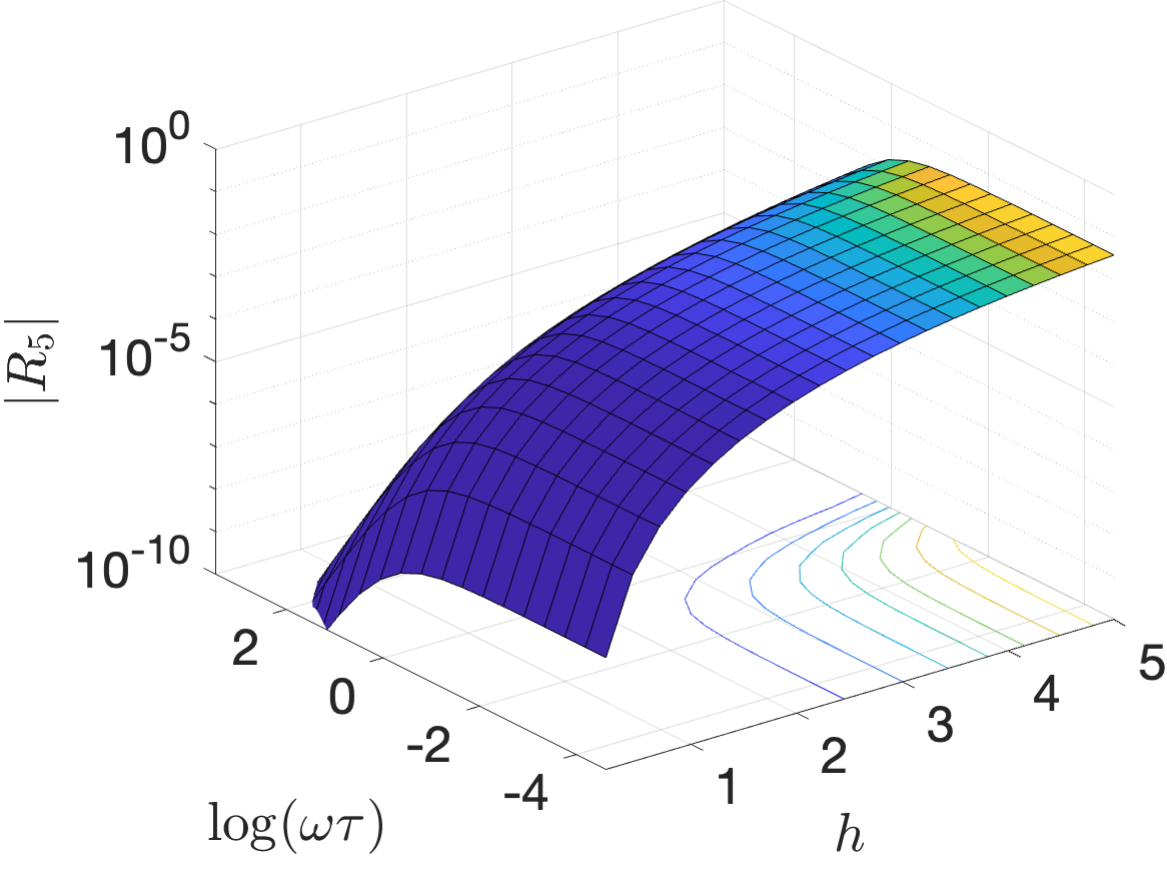}
\includegraphics[width=0.32\textwidth]{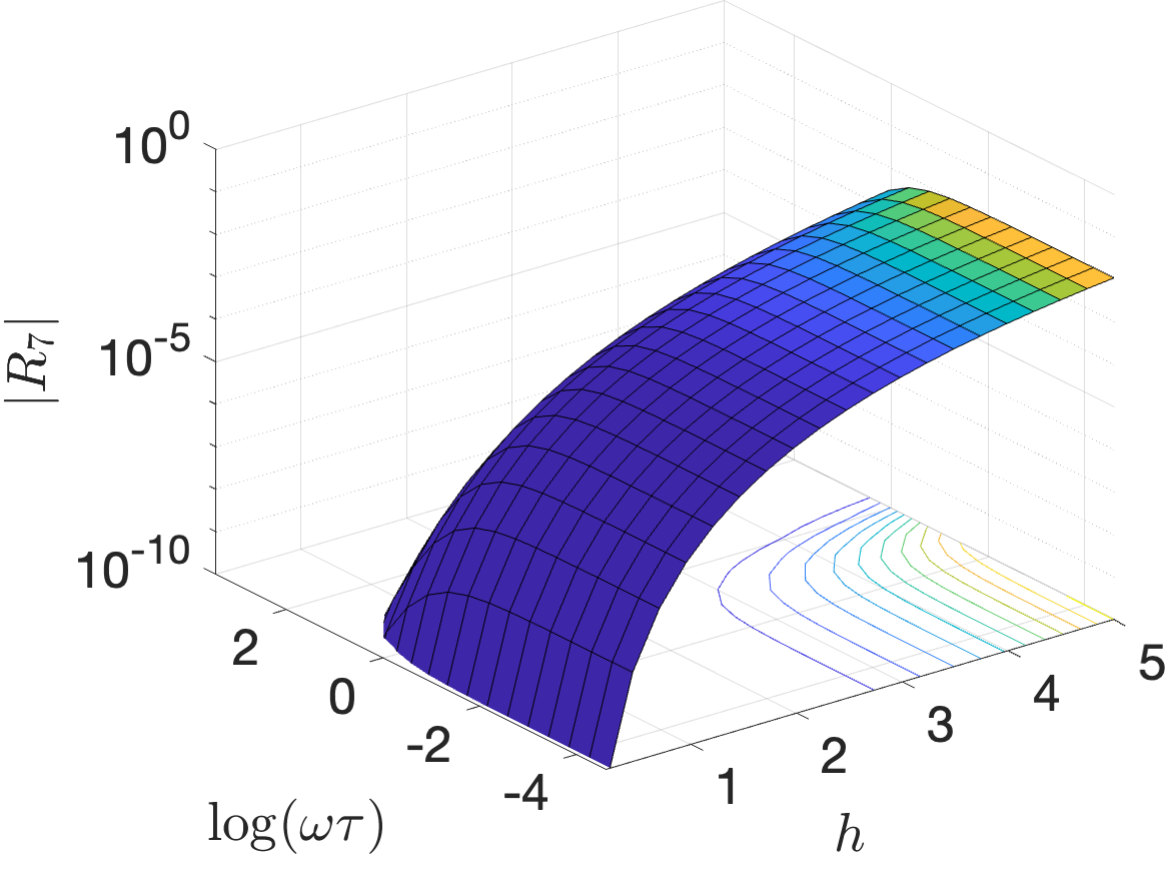}
\includegraphics[width=0.32\textwidth]{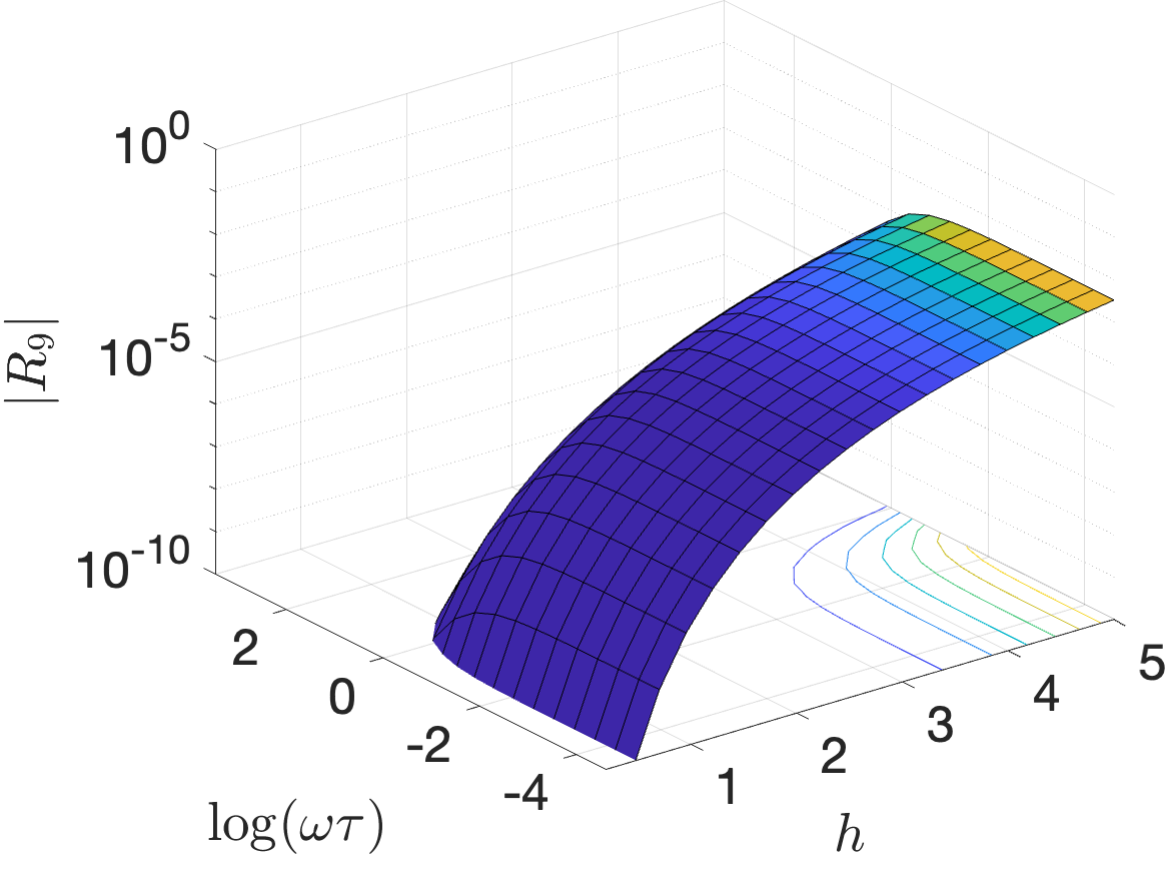}
\includegraphics[width=0.32\textwidth]{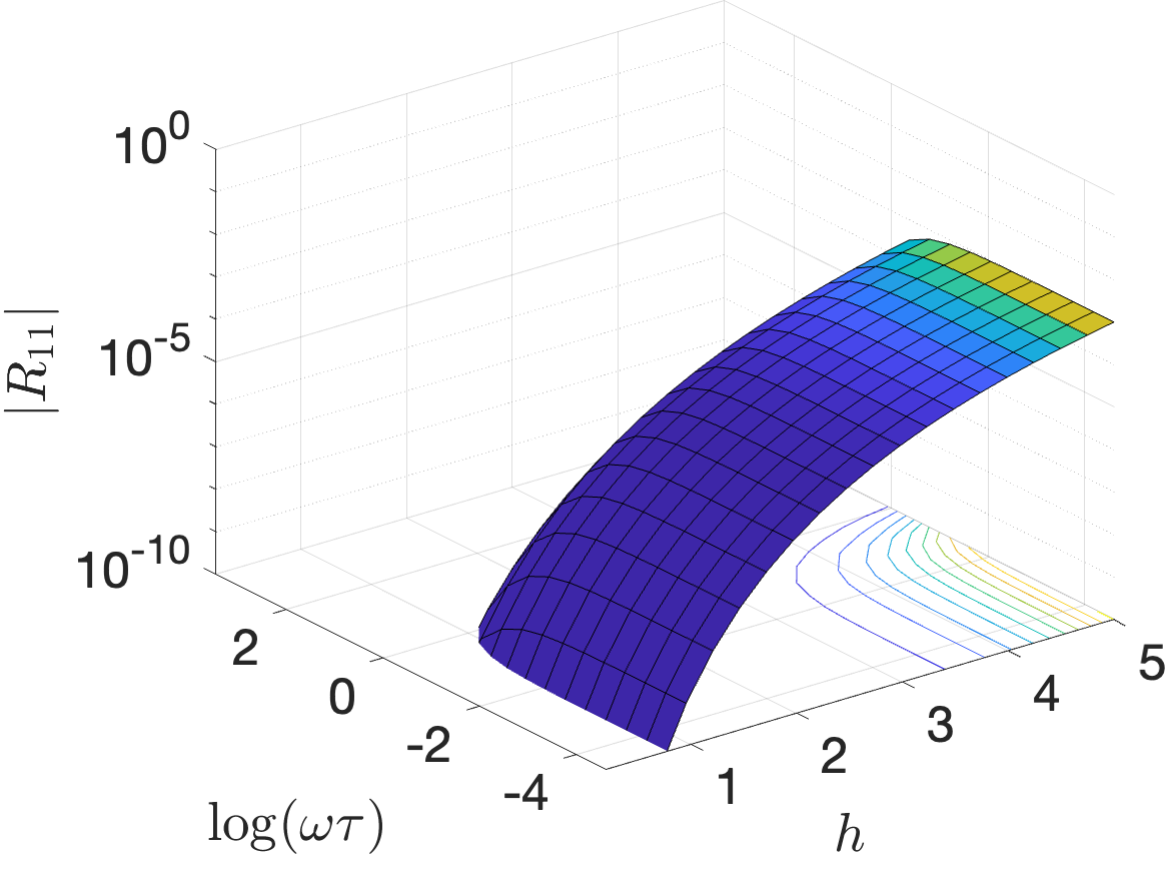}
\caption{The dimensionless magnitude $|R_n|$ of the first odd Fourier coefficients, $n=1,\ldots,11$, plotted on a logarithmic scale versus the dimensionless amplitude $h$ of the applied magnetic field and  frequency $\omega\tau$.}
\label{Rn_w_h.fig}
\end{center}
\end{figure}

Figure \ref{Rn_w_h.fig} shows the magnitude $|R_n|$ of the $n$th order harmonic response, Eq.\ \eqref{Rn_zeta}, as a function of the dimensionless amplitude $h$ of the applied field as well as the reduced frequency $\omega\tau$. 
We observe that for given field amplitude $h$ and frequency $\omega$, $|R_n|$ decreases strongly with increasing $n$.  Note the semi-logarithmic scale. 
For fixed $n$ and frequency $\omega$, on the other hand, $|R_n|$ increases with increasing $h$ since larger amplitudes promote stronger anharmonic responses. 
The increase of $|R_n|$ with $h$ is particularly strong for large $n$, reflecting the increasing importance of higher harmonics at large field amplitudes.  
Finally, for fixed $n$ and field amplitude $h$, $|R_n|$ decreases with increasing frequency. 
Such a behavior for $R_1$ is known from the Debye model where $R_1^{\rm D}\sim [1+(\omega\tau)^2]^{-1/2}$. 
The situation is more complicated for the Sh01 model, but at least the limiting cases can be discussed straightforwardly. 
For very low frequencies, $B_n$ is found to vanish (see Fig.\ \ref{AnBn_w.fig}(b)) so that $R_n$ is dominated by $A_n$. 
For large frequencies, as discussed previously, $A_n$ and $B_n$ vanish rapidly, leading to a rapid decrease of $R_n$.

\subsection{Comparisons}

Having illustrated the exact results \eqref{An_Sh01} -- \eqref{Rn_zeta} for the Sh01 model, 
a natural question is how specific or general this result is with regard  to other models for the magnetization dynamics. 
Interpreted as a phenomenological model, the magnetization equation \eqref{M_t} with \eqref{Sh01_full} can be applied to different systems with $\tau$ an effective relaxation time. 
If, on the other hand, the Sh01 model is interpreted as a simplified description for ultra-dilute (non-interacting) thermally blocked (rigid dipoles) MNPs, $\tau$ can be identified with the Brownian rotational diffusion time $\tauB$ of an individual MNP in a given solvent. In this case, the model predictions can  be compared quantitatively with the kinetic model of Martsenyuk \emph{et.\ al.}\ (MRS) \cite{MRS74}, which is generally considered a more accurate description for such systems. 
This model has been discussed extensively in the literature (see e.g.\ \cite{Odenbach_LNP763,IKH01,Deissler2014,Coffey:1993}  
and references therein) and in particular its nonlinear response to oscillating fields  
\cite{felderhof_nonlinear_2003,Yoshida2009,draack_multiparametric_2019,kuznetsov_nonlinear_2022,enpuku_harmonic_2023}. 
In the same paper  \cite{MRS74}, Martsenyuk \emph{et.\ al.}\ also introduced an effective field approximation (EFA) to arrive at a closed-form magnetization equation from the kinetic model. 

\begin{figure}[htbp]
\begin{center}
\includegraphics[width=0.47\textwidth]{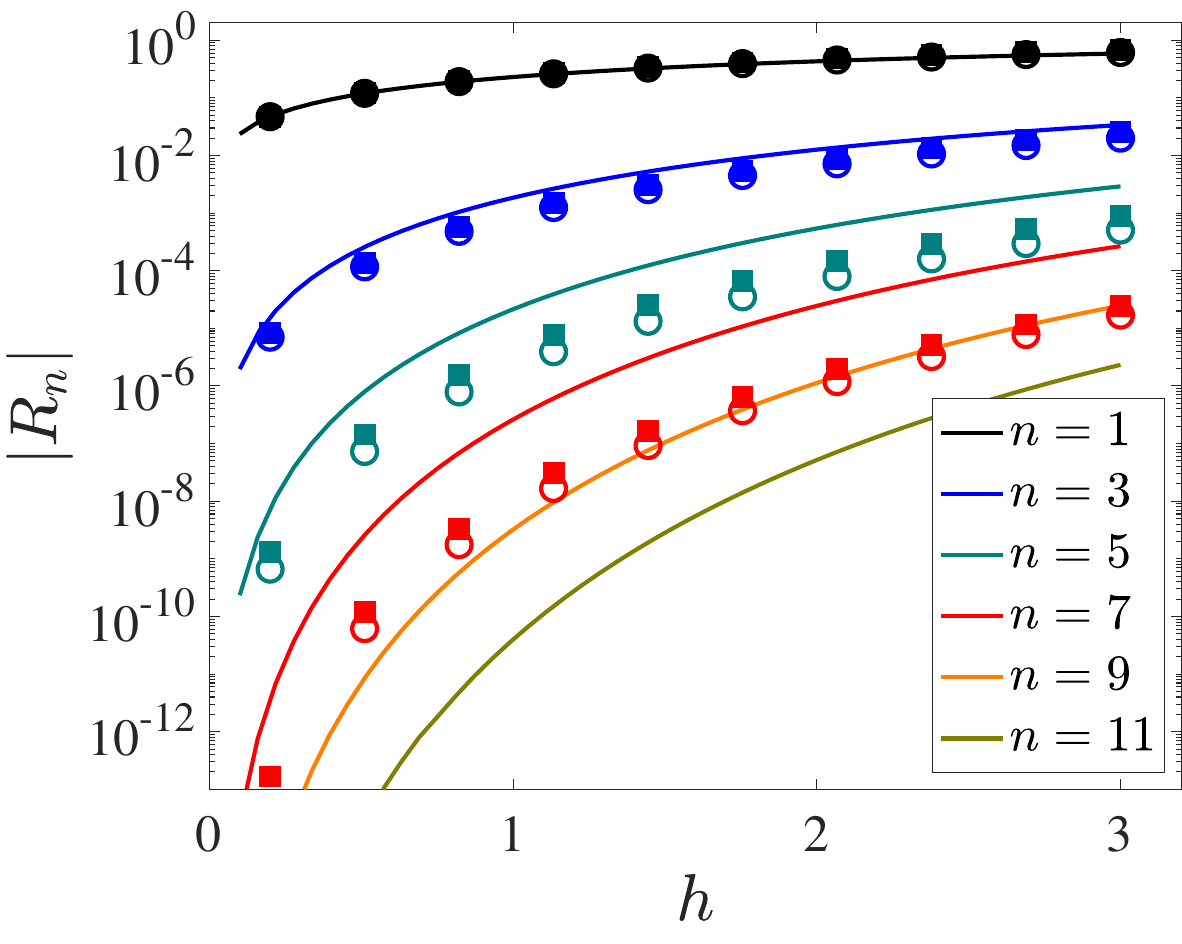}
\includegraphics[width=0.47\textwidth]{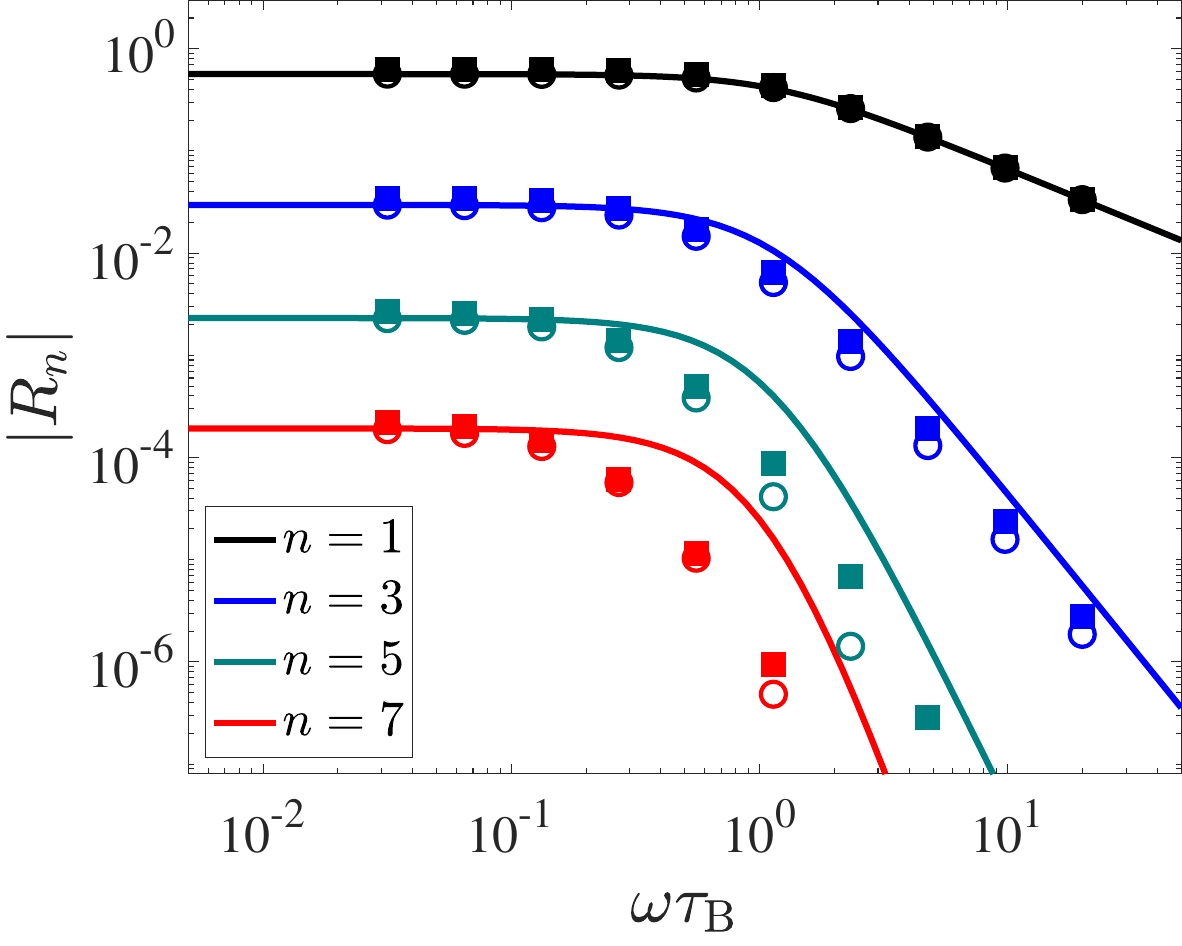}
\caption{Left: The dimensionless magnitude $|R_n|$ for Sh01 (lines), MRS  (filled symbols), and EFA (open symbols) models versus the dimensionless amplitude $h$ of the applied field.
The frequency was chosen as $\omega\tauB=1$.  
Right: Same as left panel but plotted as a function of reduced frequency $\omega\tauB$ for fixed amplitude $h=2$. 
In both panels, $n$ increases from top to bottom as indicated in the legend.}
\label{Rn_w_compare.fig}
\end{center}
\end{figure}

Figure \ref{Rn_w_compare.fig} shows a comparison of the dimensionless magnitude $|R_n|$ of the $n$th harmonic response for the Sh01, MRS, and EFA models. 
The MRS and EFA model are briefly discussed in Appendix \ref{ode.sec}. 
Accurate numerical calculations of the amplitudes of higher order responses can be challenging. 
Therefore we also provide in Appendix \ref{ode.sec} details of their numerical solution and algorithms. 
Panels (a) and (b) in Fig.\ \ref{Rn_w_compare.fig} can be seen as cross sections of Fig.\ \ref{Rn_w_h.fig} for fixed frequency and amplitude, respectively. 
On a logarithmic scale, all three models agree well for $n=1$. More pronounced differences are seen as $n$ increases. 
Interestingly, the Sh01 model overestimates whereas EFA underestimates $|R_n|$ with respect to the MRS model. 
Thereby, the EFA provides a better approximation to MRS than the Sh01 model. 
Overall, however, the qualitative behavior seen in Fig.\ \ref{Rn_w_compare.fig}  is very similar for all three models.

\begin{figure}[htbp]
\begin{center}
\includegraphics[width=0.47\textwidth]{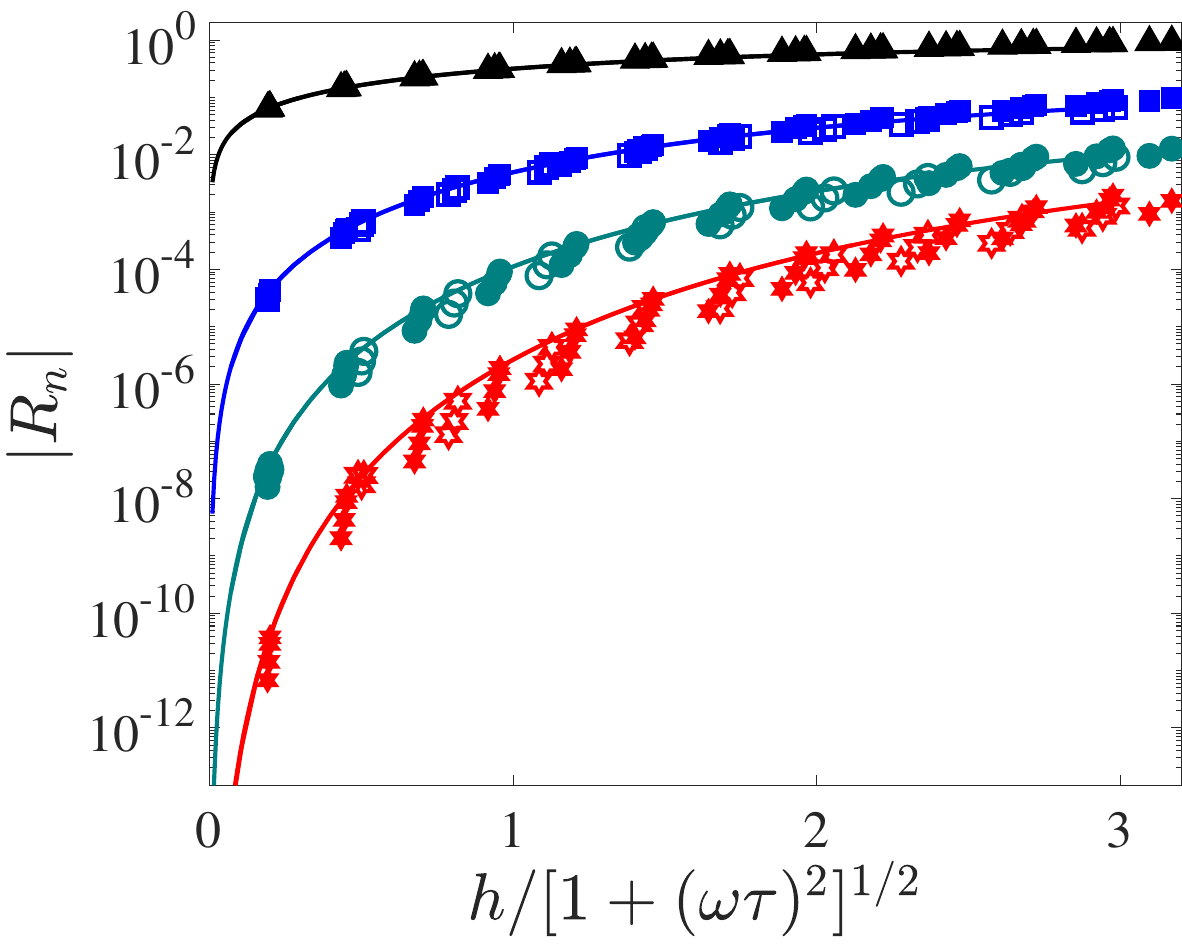}
\caption{The dimensionless magnitude $|R_n|$ for Sh01 model (lines), MRS model (filled symbols), and EFA (open symbols) versus the dimensionless scaling variable $h/\sqrt{1+(\omega\tau)^2}$.
Different amplitudes $h$ and frequencies $\omega$ have been chosen (see text). 
For a better comparison, slightly different field strengths have been chosen for MRS and EFA model. 
From top to bottom $n$ increases as $n=1, 3, 5, 7$, shown as triangles, squares, circles and diamonds, respectively.}
\label{Rn_hwscale.fig}
\end{center}
\end{figure}

A very interesting finding for the Sh01 model in Sec.\ \ref{exact.sec} is that $R_n$ depends on the field amplitude $h$ and frequency $\omega$ only via their combination,
$R_n(h,\omega\tau)=F_n(h/\sqrt{1+(\omega\tau)^2})$,
with the scaling function $F_n$ given by Eq.\ \eqref{Rn_zeta}.
In Fig.\ \ref{Rn_hwscale.fig} we show the scaling function $F_n$ for the Sh01 model for $n=1, 3, 5, 7$ as lines.
Also shown in Fig.\ \ref{Rn_hwscale.fig} are the corresponding data for $|R_{n}|$ for the MRS and EFA model obtained for equidistant field strengths $h=0.2,\ldots, 5$ and frequencies $\omega\tauB = 0.05, 0.1, 0.2, 0.3$.
While a rather good data collapse for the MRS and EFA model onto the Sh01 predictions is seen for $n=1$ and $n=3$,
the quality of data collapse becomes poorer for larger $n$ and also for larger frequencies (not shown in Fig.\ \ref{Rn_hwscale.fig}).

\begin{figure}[htbp]
\begin{center}
\includegraphics[width=0.45\textwidth]{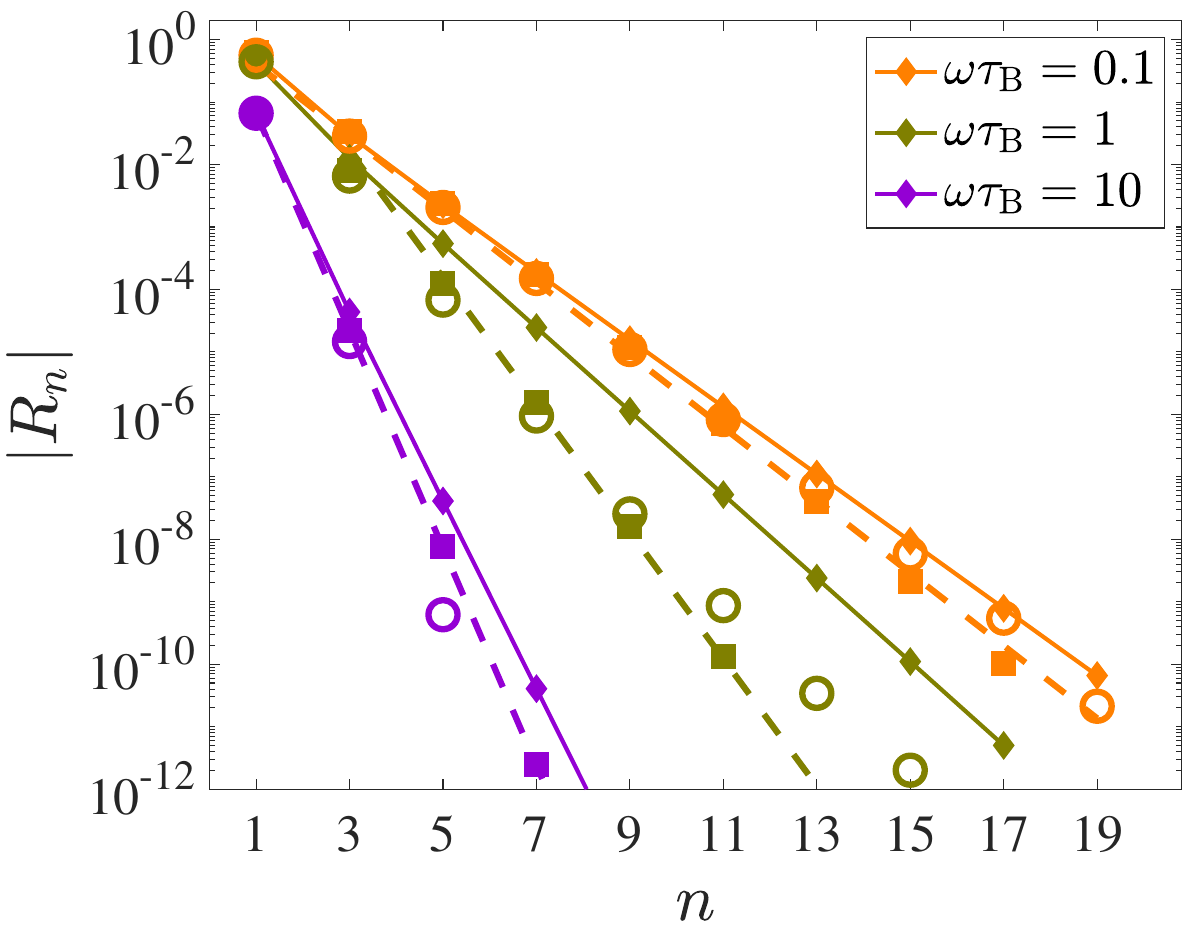}
\includegraphics[width=0.45\textwidth]{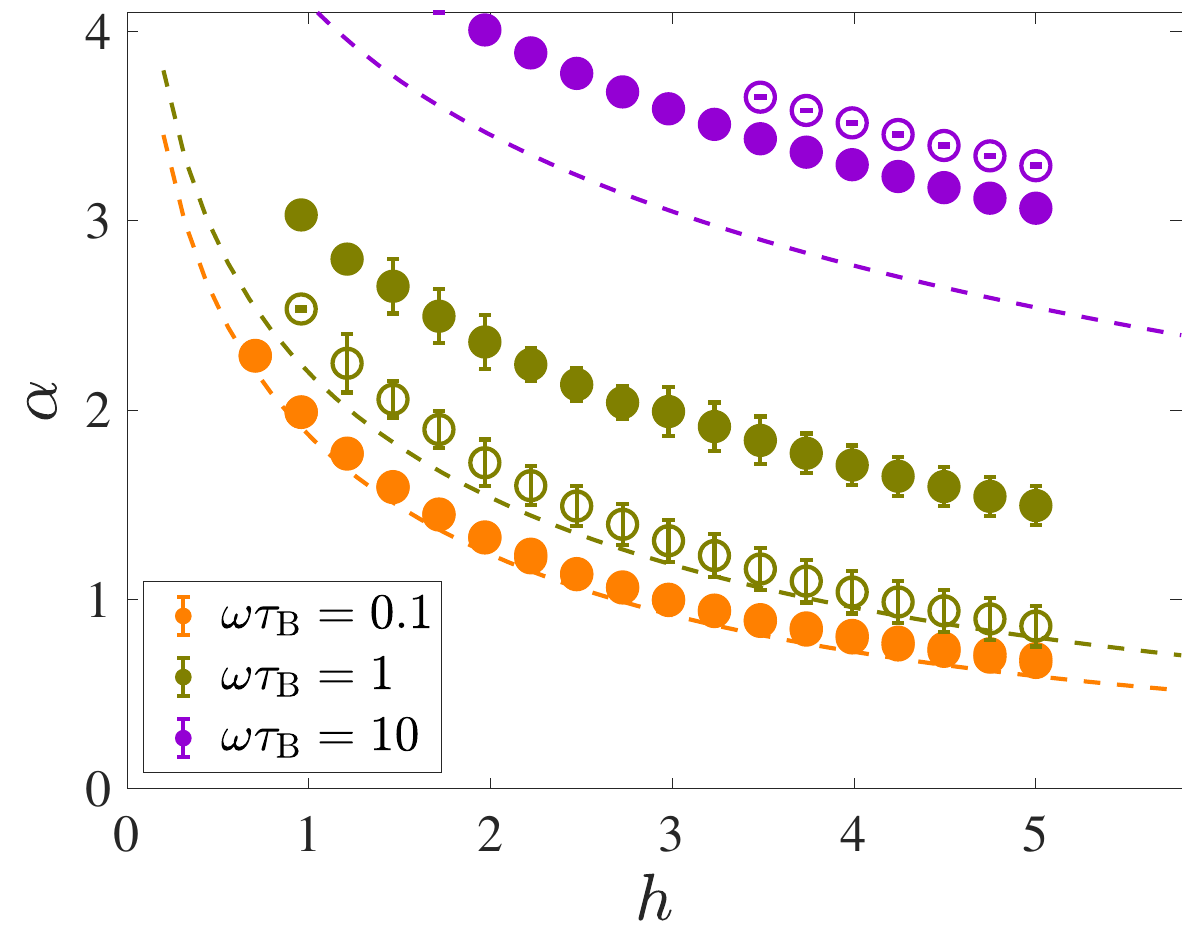}
\caption{Left: The dimensionless magnitude $|R_n|$ of $n$th harmonic response for Sh01 (diamonds), the MRS model (squares) and EFA (circles) at fixed amplitude $h=2$ for different frequencies as indicated in the legend. 
Exponential fit to MRS data for $n\geq n_{\rm min}$ (see text) are shown as dashed lines. 
Right: Filled and open circles represent the fit coefficients $\alpha$ in Eq.\ \eqref{Rn_expfit} obtained for the MRS and EFA model, respectively. Dashed lines indicate the analytical result for the Sh01 model.}
\label{Rn_n_compare.fig}
\end{center}
\end{figure}

Figure \ref{Rn_n_compare.fig}(a) shows the magnitude $|R_n|$ of the $n$th harmonic response versus $n$ for a fixed field amplitude $h=2$.
With increasing frequency of the applied field we find that $|R_n|$ decreases faster with $n$, indicating a weakening of anharmonic contributions. 
Overall we see a rapid decrease of $|R_n|$ with $n$ which makes their numerical evaluation difficult for large $n$.
From the Sh01 model, we found an exponential decay $|R_n|\sim e^{-\alpha(z) n}$ for large $n$, with
amplitude and frequency-dependent decay constant $\alpha(z) = -\ln(z/[1+\sqrt{1+z^2}])$,
see Eq.\ \eqref{Rn_scal}.
For large frequencies such as $\omega\tauB=10$, data shown in Fig.\ \ref{Rn_n_compare.fig}(a) can be described reasonably well  by a single-exponential decay. 
For smaller frequencies, however, a different decay at smaller $n$ is seen. 
We find that a double-exponential fit can give satisfactory descriptions of the data in most cases. However, uncertainties in the resulting fit parameters are often found to be very large. 
Therefore, we choose a more robust method by discarding small $n$ values so that we can fit data to 
\begin{equation} \label{Rn_expfit}
|R_n| \approx c e^{-\alpha n} \quad\text{for}\ n\geq n_{\rm min} , 
\end{equation}
where $n_{\rm min}$ denotes the onset of the exponential decay for large $n$. 
The values of $\alpha$ obtained from fits of Eq.\ \eqref{Rn_expfit} to the MRS and EFA data are shown in Fig.\ \ref{Rn_n_compare.fig}(b) as filled and open symbols, respectively.
Guided by the results for the Sh01 model, we choose $n_{\rm min}=7$ for $\omega\tauB=0.1$ and $1$, and 
 $n_{\rm min}=1$ for $\omega\tauB=10$.

\section{Discussion and Conclusions}

While the magnetization response to weak oscillating fields has been intensively studied in the literature, much less is known for the case where large-amplitude oscillatory fields are applied. The latter play an important role in several biomedical applications, but little guidance from theory is available so far.

Here, we provide the analytic solution of the fully nonlinear magnetization response to oscillatory fields valid for a particular model of the magnetization dynamics (the Sh01 model) and for not too large amplitudes.
The exact solution shows that the nonlinear response obeys a number of remarkable properties, including symmetries and sign changes of the in-phase and out-of-phase contributions as well as a scaling behavior of the magnitude of the response in terms of a combination of field amplitude and frequency.
The scaling variable is also found to govern the exponential decrease of the magnitude of the response with increasing order of harmonics.

A comparison of these analytical results to numerical solutions of rival models for the magnetization dynamics of non-interacting, thermally blocked MNPs (MRS and EFA models) gives an indication how specific or general the exact result obtained for the Sh01 model is.
We find that the MRS and EFA models show the same qualitative features of the nonlinear response  found in the Sh01 model. This is a strong indication that the symmetries and number of zero-crossings of the
in-phase and out-of-phase contributions to the nonlinear response indeed hold more generally.
Differences between the models are seen in the quantitative values of the response, with relative deviations becoming more pronounced for higher harmonics and for larger amplitudes and frequencies. 
Similar comparisons between the Sh01, MRS and EFA models have already been discussed for the third order susceptibility, where rather good quantitative agreement was found except near $\omega\tauB\sim 1$  \cite{ilg_medium_2024}. 
Here, we also find that the scaling relation found for the magnitude of the nonlinear response in the Sh01 model also holds approximately  for the first seven harmonics in MRS and EFA model for not too large frequencies.
Therefore, in this regime, the nonlinear response can be described in terms of a single scaling variable that combines the field amplitude and frequency.
Finally, with the exception of the first few harmonics, all three models show an exponential decay of the magnitude of the nonlinear response $|R_n|$ with harmonic number $n$.
The rate of this exponential decay is well described for all three models by the analytical result for the Sh01 model for low enough frequencies, whereas quantitative differences appear for larger frequencies.

Strong perturbations -- like the large-amplitude oscillatory magnetic fields considered here -- probe details of nonlinearities inherent in model systems.
The resulting nonlinear response is therefore typically non-universal and may depend on details of the models.
Identifying common properties in the nonlinear response of different models is not only highly interesting from a theoretical point of view, but also very useful in practice where it is not always straightforward to decide on the appropriate model formulation. 
Future work will need to include the influence of internal Néel relaxation \cite{Coffey:1993,Deissler2014,ilg:DJ2023} and extend studies 
towards biological environments where clusters of MNPs are frequently encountered \cite{moor_particle_2022,egler-kemmerer_real-time_2022}.
It will be interesting to see how the present results compare when these additional effects are included.

\section*{Acknowledgments}

Stimulating discussions with Prof.\ Inge Herrmann, ETH Zürich,
 are gratefully acknowledged.

\begin{appendix}

\section{Convergence of analytic result}
The analytical result in Eqs.\ \eqref{An_Sh01} and \eqref{Bn_Sh01} for the Sh01 model is given in terms of an infinite sum \eqref{Rn_zeta} involving binomial coefficients and the Rieman zeta function. 
We found this formulation not particularly well suited for numerical evaluations. 
Instead, the equivalent equation 
\begin{equation}
R_n(h,y) = \sum_{k=0}^\infty 
\frac{4 \BB_{2k+n+1}}{(2k+n+1)(k!)(k+n)!} 
\left( \frac{h}{\sqrt{1+y^2}} \right)^{n+2k}
\label{Rn_Bernoulli}
\end{equation}
was found to be preferable when evaluated in Matlab\texttrademark\ with in-built Bernoulli numbers $\BB_j$. 

\begin{figure}[htbp]
\begin{center}
\includegraphics[width=0.4\textwidth]{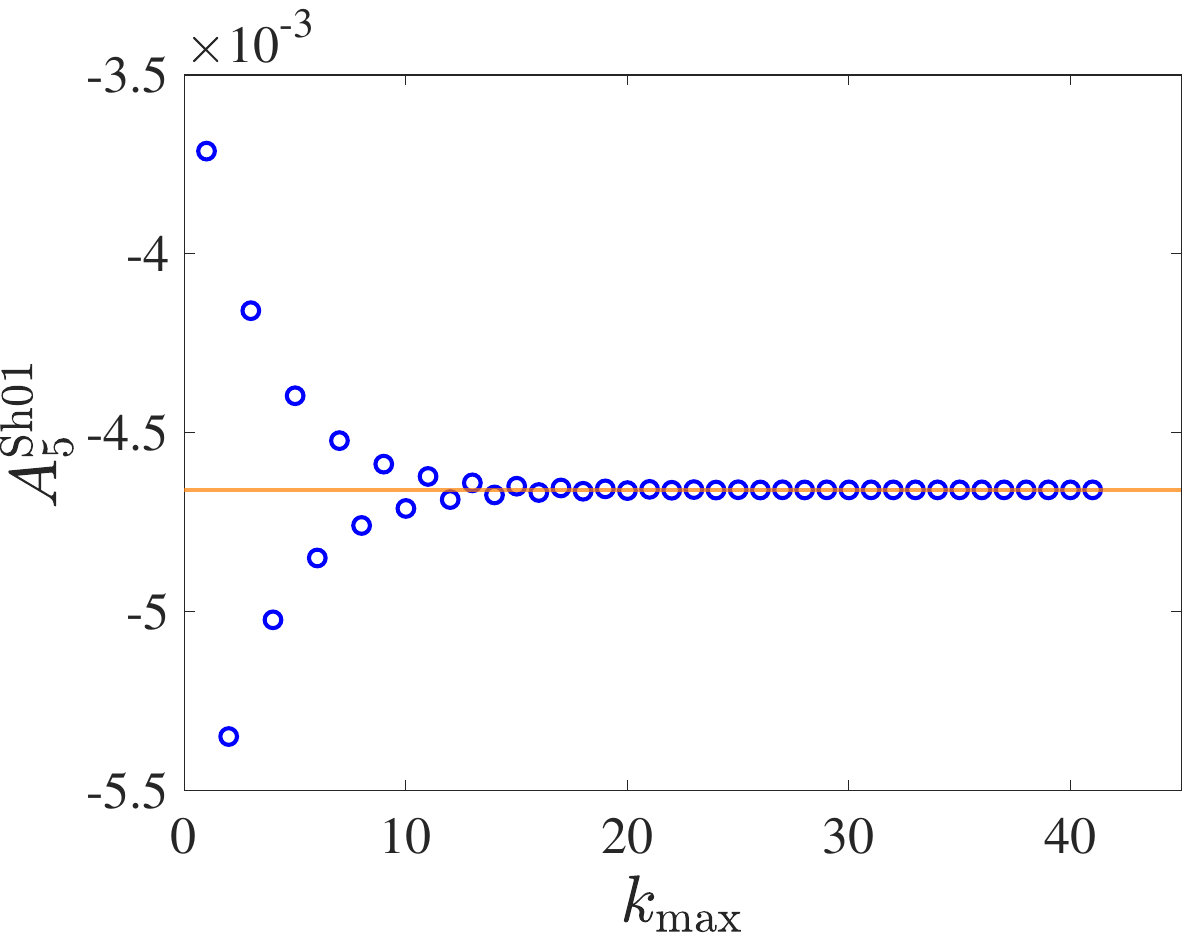}
\includegraphics[width=0.4\textwidth]{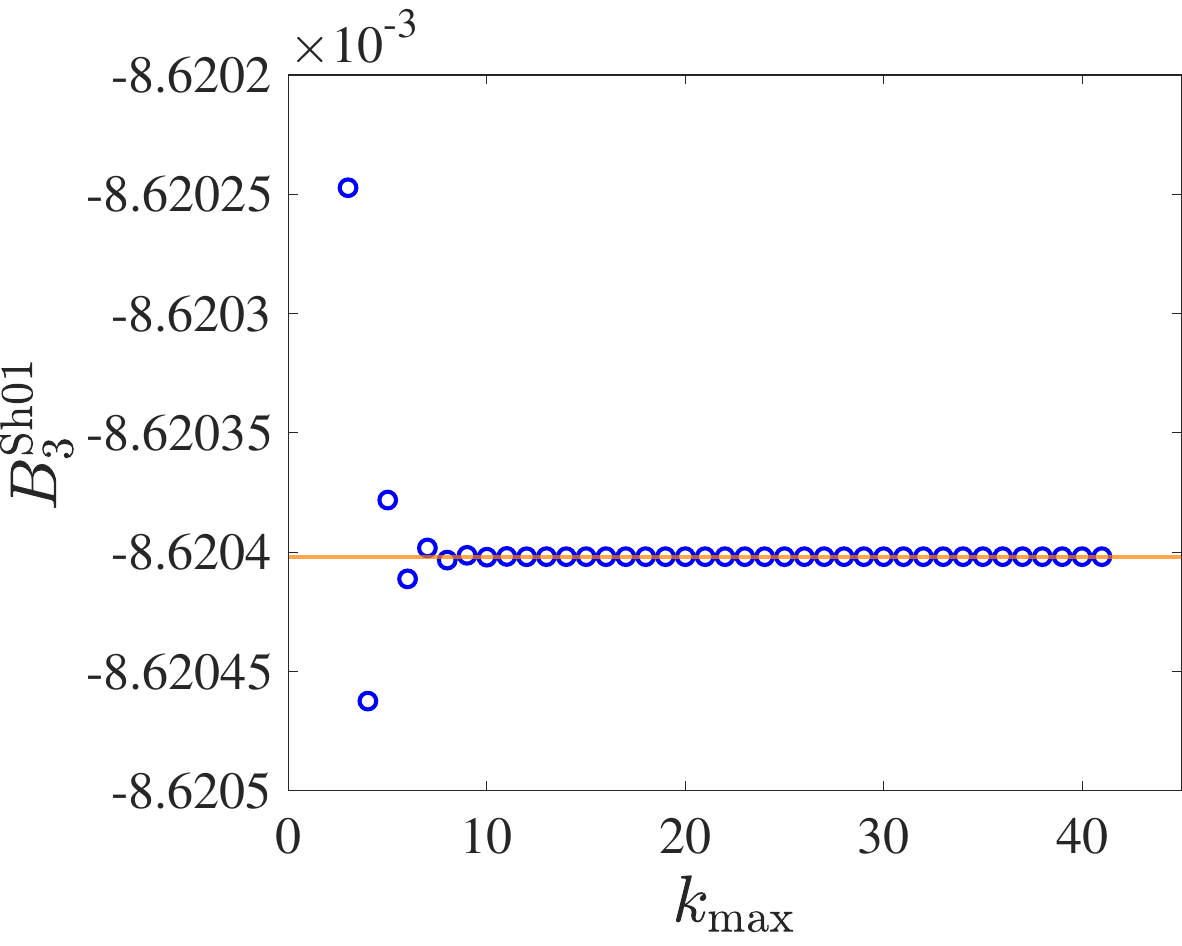}
\caption{Convergence of $A_5^{\rm Sh01}$ from Eq.\ \eqref{Rn_Bernoulli} for $h=3$, $\omega\tau=0.5$ (left) and $B_3^{\rm Sh01}$ for $h=2$, $\omega\tau=0.1$ (right) when the infinite sum is truncated at $k_{\rm max}$. The horizontal line shows the result of a high-precision numerical evaluation of the corresponding integrals (see text).
}
\label{An_converge.fig}
\end{center}
\end{figure}

In practice we choose a tolerance level $\varepsilon$ and truncate the sum \eqref{Rn_Bernoulli} when the absolute magnitude of term $k$ is smaller than $\varepsilon$. Typically we choose $\varepsilon=10^{-12}$. 
Convergence is found to be faster the smaller $n$ and $h$ is and the larger $\omega$. 
Figure \ref{An_converge.fig}(a) and (b) illustrate the convergence of $A_5$ and $B_3$, respectively, from Eqs.\ \eqref{An_Sh01} and \eqref{Bn_Sh01} with Eq.\ \eqref{Rn_Bernoulli} for two representative choices of parameters. 
As a further check, the results of the summation are also compared to accurate numerical evaluations (with absolute tolerance $10^{-12}$) of the integrals \eqref{An_integral} with the exact solution 
$M(t)=M(\xie(t))$ where $\xie(t)$ is obtained from Eq.\ \eqref{He_sol}. 

\section{Numerical Fourier Transformation} \label{FT.sec}

Fast Fourier Transform (FFT) is a highly versatile and very frequently used method to efficiently determine the frequency content of time series. However, standard FFT routines come with some limitations.  
Issues of FFT with distortion of lower-amplitude peaks are particularly problematic for situations considered here and corresponding applications such as MPS.
Therefore, much more precise versions of numerical Fourier transforms have been developed recently \cite{henry_ultra-precise_2023}.

Here, we benefit greatly from the fact that the frequencies $n\omega$ are known exactly.
Therefore, the remaining task is to evaluate the Fourier integrals \eqref{An_integral} to high precision.
To achieve this, we use an implementation of Simpson's rule of integration in Matlab\texttrademark\ with global adaptive quadrature with $10^4$ steps per period $T_\omega$ and an error tolerance of $\epsilon=10^{-12}$.

Various tests of the algorithm and our implementation were performed.
First, input data for the Fourier integrals were generated for several choices of field amplitude and frequencies by numerically evaluating Eqs.\ \eqref{M_t} and \eqref{He_sol}.
Numerical issues in evaluating the Langevin function $L(\xie(t))$ for $|\xie(t)|\ll 1$ were solved by switching to a fifth-order expansion $L(x)\approx x/3 - x^3/45 + 2x^5/945 + {\cal O}(x^7)$ for $|x|<10^{-2}$.
Our numerical results for the integrals \eqref{An_integral} show very good agreement with the analytical results \eqref{An_Sh01} -- \eqref{Rn_zeta} down to values of $|A_n|, |B_n| \sim 10^{-10}$, whereas with standard FFT routines we found noticeable deviations already around $10^{-6}$.

\section{Numerical Solution to Time Evolution Equations} \label{ode.sec}

In view of analyzing other models of the magnetization dynamics for which exact solutions are not available, we also consider obtaining numerical data for $M(t)$ by solving Eq.\ \eqref{Sh01_full} numerically.
In the present case, the ordinary differential equation (ODE) simplifies to
$\tau \frac{\dd m}{\dd t}= -L'(\xie(t))[\xie(t) - h(t)]$, where $m(t)=M(t)/\Msat$ and $L'(\xie(t))=\frac{\dd L(x)}{\dd x}|_{x=\xie(t)}$ denotes the derivative of the Langevin function evaluated at $\xie(t)$.
To achieve a comparable accuracy of at least $10^{-10}$ in the resulting Fourier integrals, we use a high-order Runge-Kutta method (\texttt{ode89} in Matlab\texttrademark) with tolerance $10^{-8}$.
To be sure to eliminate also tiny signs of initial transients, we solve the ODE for $50$ oscillation periods and only use the last period as input data for Eq.\ \eqref{An_integral}.
With this method, data in Fig.\ \ref{Rn_n_compare.fig} are indistinguishable from the exact results on this scale. 

The magnetization equation within EFA proposed in Ref.\ \cite{MRS74} can be written as 
$\tauB \frac{\dd M}{\dd t}= -[1 - h(t)/\xie(t)]M(t)$, where again $M(t)=\Msat L(\xie(t))$ is given in terms of the dimensionless effective field $\xie(t)$. 
Closed-form solutions to this model are not known in general. 
We obtain numerical solutions for the EFA model using the same algorithms for solving the ODE and subsequent Fourier analysis as described previously. To avoid the use of the inverse Langevin function, it is more convenient to numerically solve the corresponding ODE for the effective field which reads 
$\tauB \frac{\dd \xie}{\dd t}= -[1 - h(t)/\xie(t)]L(\xie(t))/L'(\xie(t))$, and subsequently calculate the magnetization.

For the kinetic MRS model  we follow Ref.\ \cite{Coffey:1993,felderhof_nonlinear_2003} and use an expansion of the probability density function in Legendre polynomials $P_\ell(x)$, $f(\be;t)=\sum_{\ell=0}^{\infty}c_\ell(t)P_\ell(\be\cdot\hat{\bH})$, from which
a coupled system of ODEs for the expansion coefficients $c_\ell$ can be obtained \cite{Coffey:1993}, 
\begin{equation} \label{dtcn_MRS}
 \frac{2\tauB}{\ell(\ell+1)} \frac{\dd c_\ell}{\dd t} =  h(t)\left[\frac{c_{\ell-1}}{2\ell-1} - \frac{c_{\ell+1}}{2\ell+3}\right] - c_\ell ,
\end{equation}
with constant $c_0=1/(4\pi)$ ensuring correct normalization of the probability density for all times.
For numerical solutions, the infinite system needs to be truncated at a certain $\ell_{\rm max}$, corresponding to
the number of Legendre polynomials needed  to achieve a prescribed accuracy. 
While $\ell_{\rm max}=30$ was used for $h=20$ \cite{Yoshida2009}, we found $\ell_{\rm max}=11$ to be sufficient for 
 field amplitudes $h<\pi$ investigated here, but mostly used $\ell_{\rm max}=20$ which gives indistinguishable results on the scales used in the figures.  
The resulting system of coupled ODEs is then integrated numerically with the same method as used for the other models. 
Also the very same method for the Fourier analysis is used as previously.

Data for the MRS model shown here are obtained with an alternative and, in principle, more accurate method. 
We note that on the scale shown in the previous figures, results are indistinguishable from the previously described method. 
Instead of numerically solving the system of ODEs \eqref{dtcn_MRS} and subsequent Fourier transformation once initial transients have decayed, we can use the ansatz 
\begin{equation} \label{cell_expansion}
c_\ell(t) = \sum_{n=1}^\infty \big( a_{\ell,n}\cos(n\omega t) + b_{\ell,n}\sin(n\omega t) \big)
\end{equation}
to arrive at a linear system of equations for the coefficients $a_{\ell,n}, b_{\ell,n}$, 
\begin{align} \label{a_elln}
\frac{2n\omega\tauB}{\ell(\ell+1)} b_{\ell,n} + a_{\ell,n} & = \frac{h}{2}\left( 
\frac{a_{\ell-1,n-1}+a_{\ell-1,n+1}}{2\ell-1} - \frac{a_{\ell+1,n-1}+a_{\ell+1,n+1}}{2\ell+3}
\right) ,\\
-\frac{2n\omega\tauB}{\ell(\ell+1)} a_{\ell,n} + b_{\ell,n} & = \frac{h}{2}\left( 
\frac{b_{\ell-1,n-1}+b_{\ell-1,n+1}}{2\ell-1} - \frac{b_{\ell+1,n-1}+b_{\ell+1,n+1}}{2\ell+3}
\right) , 
\label{b_elln}
\end{align}
for $\ell\geq 2$, while for $\ell=1$ the corresponding equations read 
\begin{align}
n\omega\tauB  b_{1,n} + a_{1,n} & = c_0 h \delta_{n,1} - \frac{h}{10} \left( a_{2,n-1} + a_{2,n+1} \right) , \\
-n\omega\tauB  a_{1,n} + b_{1,n} & = - \frac{h}{10} \left( b_{2,n-1} + b_{2,n+1} \right) . 
\label{b_1n}
\end{align}
To solve the infinite system of equations \eqref{a_elln} -- \eqref{b_1n} for given amplitude $h$ and frequency $\omega$ of the applied field, we truncate the expansion in Legendre polynomials at some maximum degree $\ell_{\rm max}$ as discussed. In addition,  the expansion in terms of higher harmonics \eqref{cell_expansion} needs to be truncated at some maximum order $n_{\rm max}$. 
The truncated system can be solved to find the $2\ell_{\rm max}n_{\rm max}$ coefficients $\{a_{\ell,n}, b_{\ell,n}\}$. 
Here, we are only interested in the magnetization response \eqref{M_Fourier}, for which the corresponding coefficients are obtained by  
$A_n = (4\pi/3)a_{1,n}$ and $B_n = (4\pi/3)b_{1,n}$.

\begin{figure}[htbp]
\begin{center}
\includegraphics[width=0.4\textwidth]{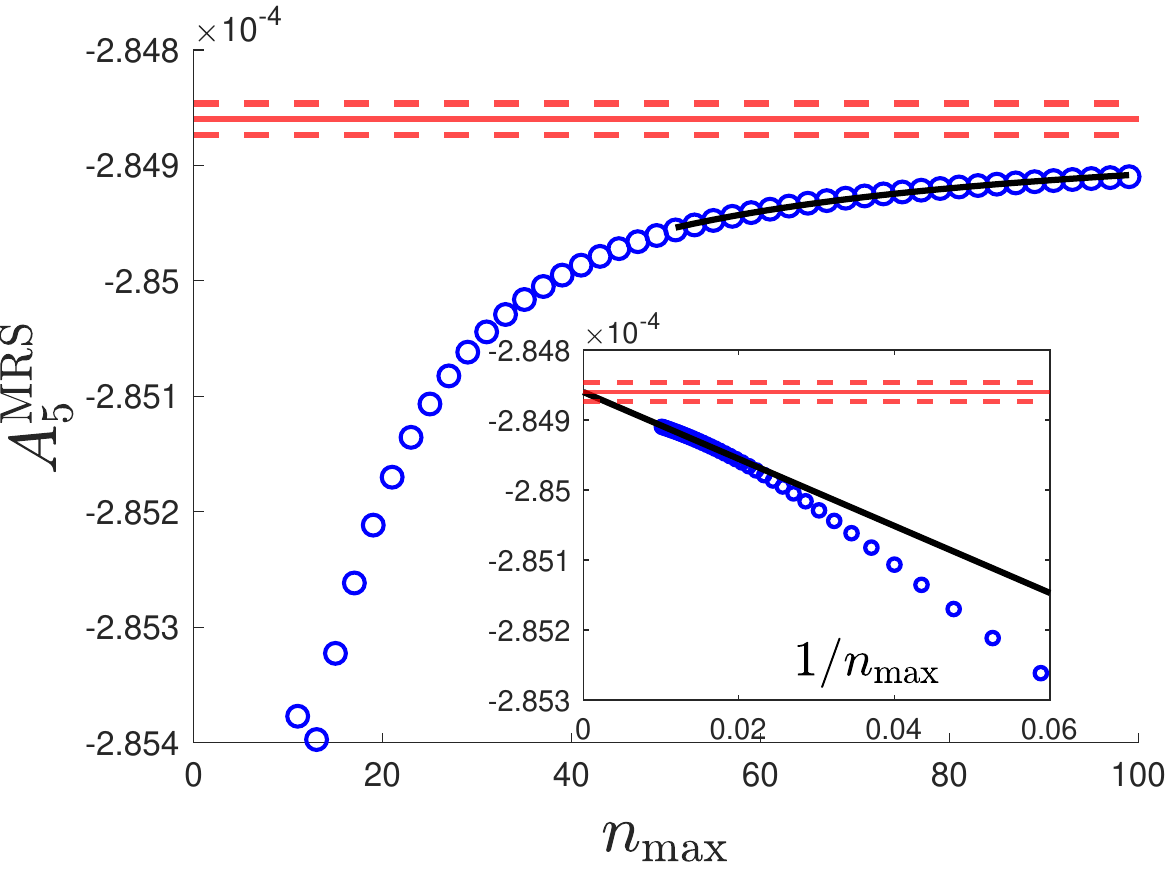}
\includegraphics[width=0.4\textwidth]{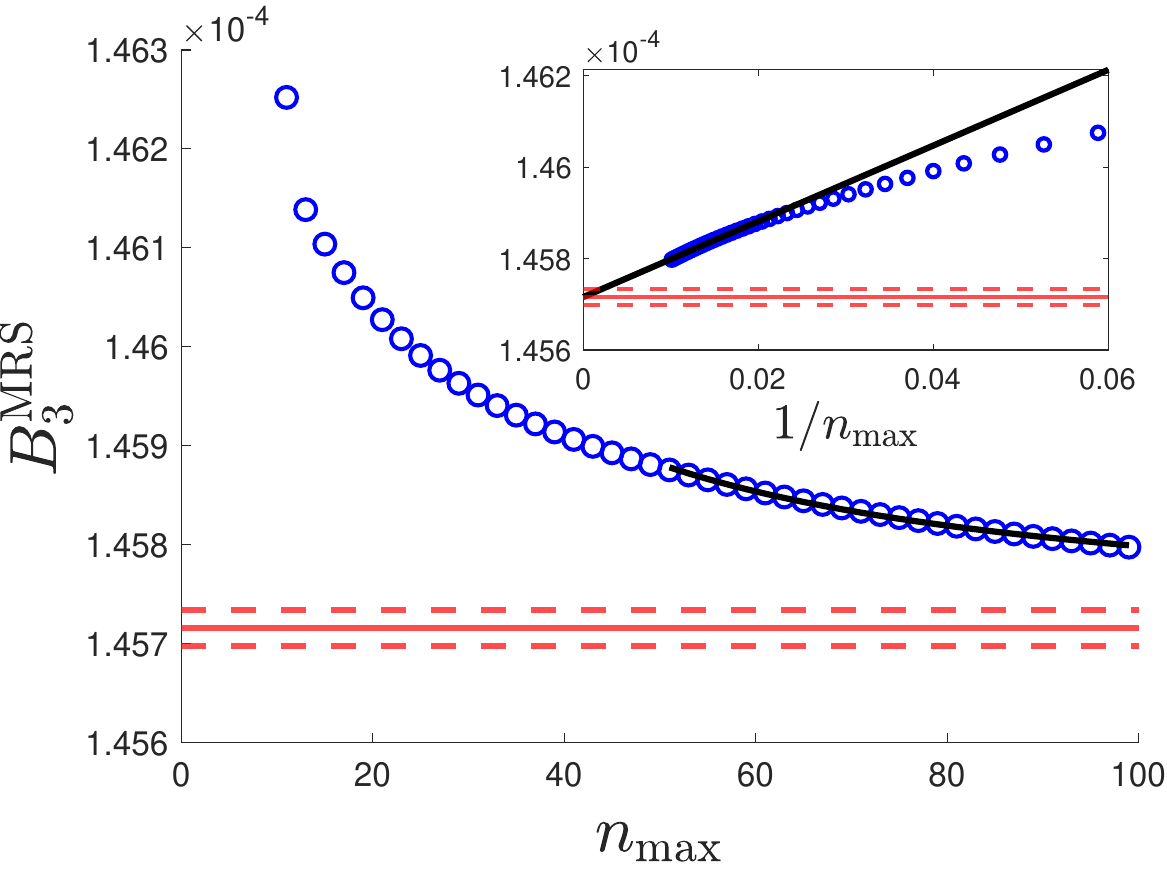}
\caption{Convergence of $A_5^{\rm MRS}$ from solution to Eqs.\ \eqref{a_elln} -- \eqref{b_1n}  for $h=3$, $\omega\tauB=0.5$ 
(left) and $B_3^{\rm MRS}$ for $h=2$, $\omega\tau=0.1$ (right) 
when the infinite system is truncated at $\ell_{\rm max}=20$ and varying $n_{\rm max}$. 
Solid black lines indicate  linear fits to $1/n_{\rm max}$ as described in the text. 
Red horizontal lines show the corresponding results of extrapolation to $n_{\rm max}\to\infty$ with dashed lines the confidence interval. 
The insets show the same data but plotted versus $n_{\rm max}^{-1}$. 
}
\label{An_converge_MRS.fig}
\end{center}
\end{figure}

We implemented truncated Eqs.\ \eqref{a_elln} -- \eqref{b_1n}  in Matlab and used its efficient algorithm to solve the resulting linear system for $a_{\ell,n}, b_{\ell,n}$. 
As observed  for the numerical solution to the corresponding ODE, we find results to converge quickly  
 with $\ell_{\rm max}$ and choose $\ell_{\rm max}=20$. Increasing $\ell_{\rm max}$ gives indistinguishable results for field amplitudes considered here. 
On the contrary, we find results to converge very slowly with increasing $n_{\rm max}$. 
As an example, Fig.\ \ref{An_converge_MRS.fig}(a) and (b) show the result for $A_5$ and $B_3$, respectively, obtained by solving Eqs.\ \eqref{a_elln} -- \eqref{b_1n} with $\ell_{\rm max}=20$, $h=3$, $\omega\tauB=0.5$  for different choices of $n_{\rm max}$. 
We see from Fig.\ \ref{An_converge_MRS.fig} that the results have not properly converged for $n_{\rm max}=100$, even looking at the third and fifth harmonics only. 
One explanation could be that the right-hand side of Eqs.\ \eqref{a_elln}, \eqref{b_elln} is proportional to $1/\ell$, promoting fast convergence with respect to the maximum degree $\ell_{\rm max}$ of Legendre polynomials. 
However, no corresponding decrease with order $n$ of harmonics is present. 
Since increasing $n_{\rm max}$ much further becomes computationally inefficient, we use instead the 
fit functions $X_n(n_{\rm max}) = X_n^\infty( 1 - K_n/n_{\rm max})$ with constant $K_n$ for $X\in\{A,B\}$ and  $n_{\rm max}>50$, to approximate the approach to the limit values $A_n^\infty$ and $B_n^\infty$. 
Horizontal red lines show $A_5^\infty$ and $B_3^\infty$ obtained in this way together with the corresponding confidence intervals obtained from uncertainties in the fitted values.  
\end{appendix}



\begin{thebibliography}{10}

\bibitem{Gaspard_irreversible2022}
Pierre Gaspard.
\newblock {\em The Statistical Mechanics of Irreversible Phenomena}.
\newblock Cambridge University Press, Cambridge, 2022.

\bibitem{Odenbach_LNP763}
S.~Odenbach, editor.
\newblock {\em Colloidal Magnetic Fluids}, volume 763 of {\em Lecture Notes in
  Phys.}
\newblock Springer, Berlin, 2009.

\bibitem{Ferguson:2013gi}
R~Matthew Ferguson, Amit~P Khandhar, Christian Jonasson, Jakob Blomgren,
  Christer Johansson, and Kannan~M Krishnan.
\newblock {Size-Dependent Relaxation Properties of Monodisperse Magnetite
  Nanoparticles Measured Over Seven Decades of Frequency by AC Susceptometry}.
\newblock {\em IEEE Transactions on Magnetics}, 49(7):3441--3444, July 2013.

\bibitem{kole_engineering_2021}
Madhusree Kole and Sameer Khandekar.
\newblock Engineering applications of ferrofluids: A review.
\newblock {\em Journal of Magnetism and Magnetic Materials}, 537:168222, 2021.

\bibitem{rivera-rodriguez_emerging_2021}
Angelie Rivera-Rodriguez and Carlos~M. Rinaldi-Ramos.
\newblock Emerging biomedical applications based on the response of magnetic
  nanoparticles to time-varying magnetic fields.
\newblock {\em Annual Review of Chemical and Biomolecular Engineering},
  12(1):163--185, 2021.

\bibitem{ilg_medium_2024}
Patrick Ilg.
\newblock Medium amplitude field susceptometry ({MAFS}) for magnetic
  nanoparticles.
\newblock {\em Journal of Magnetism and Magnetic Materials}, 610:172540, 2024.

\bibitem{ilg_stochastic_2024}
Patrick Ilg.
\newblock Stochastic thermodynamics and fluctuations in heat released by
  magnetic nanoparticles in response to time-varying fields.
\newblock {\em Physical Review B}, 109(17):174301, 2024.

\bibitem{wu_magnetic_2020}
Kai Wu, Diqing Su, Renata Saha, Jinming Liu, Vinit~Kumar Chugh, and Jian-Ping
  Wang.
\newblock Magnetic particle spectroscopy: A short review of applications using
  magnetic nanoparticles.
\newblock {\em {ACS} Applied Nano Materials}, 3(6):4972--4989, 2020.

\bibitem{yang_applications_2022}
Xue Yang, Guoqing Shao, Yanyan Zhang, Wei Wang, Yu~Qi, Shuai Han, and Hongjun
  Li.
\newblock Applications of magnetic particle imaging in biomedicine:
  Advancements and prospects.
\newblock {\em Frontiers in Physiology}, 13:898426, 2022.

\bibitem{harvell-smith_magnetic_2022}
Stanley Harvell-Smith, Le~Duc Tung, and Nguyen Thi~Kim Thanh.
\newblock Magnetic particle imaging: tracer development and the biomedical
  applications of a radiation-free, sensitive, and quantitative imaging
  modality.
\newblock {\em Nanoscale}, 14(10):3658--3697, 2022.

\bibitem{vogel_critical_2022}
Patrick Vogel, Martin~Andreas R{\"u}ckert, Bernhard Friedrich, Rainer Tietze,
  Stefan Lyer, Thomas Kampf, Thomas Hennig, Lars D{\"o}lken, Christoph Alexiou,
  and Volker~Christian Behr.
\newblock Critical offset magnetic {PArticle} {SpectroScopy} for rapid and
  highly sensitive medical point-of-care diagnostics.
\newblock {\em Nature Communications}, 13(1):7230, 2022.

\bibitem{moor_particle_2022}
Lorena Moor, Subas Scheibler, Lukas Gerken, Konrad Scheffler, Florian Thieben,
  Tobias Knopp, Inge~K. Herrmann, and Fabian H.~L. Starsich.
\newblock Particle interactions and their effect on magnetic particle
  spectroscopy and imaging.
\newblock {\em Nanoscale}, 14(19):7163--7173, 2022.

\bibitem{felderhof_nonlinear_2003}
B~U Felderhof and R~B Jones.
\newblock Nonlinear response of a dipolar system with rotational diffusion to
  an oscillating field.
\newblock {\em J. Phys.: Condens. Matter}, 15(15):S1363--S1378, 2003.

\bibitem{Yoshida2009}
T.~Yoshida and K.~Enpuku.
\newblock Simulation and quantitative clarification of ac susceptibility of
  magnetic fluid in nonlinear brownian relaxation region.
\newblock {\em Jpn. J. Appl. Phys.}, 48:127002, 2009.

\bibitem{kuznetsov_nonlinear_2022}
Andrey~A. Kuznetsov and Alexander~F. Pshenichnikov.
\newblock Nonlinear response of a dilute ferrofluid to an alternating magnetic
  field.
\newblock {\em Journal of Molecular Liquids}, 346:117449, 2022.

\bibitem{draack_multiparametric_2019}
Sebastian Draack, Niklas Lucht, Hilke Remmer, Michael Martens, Birgit Fischer,
  Meinhard Schilling, Frank Ludwig, and Thilo Viereck.
\newblock Multiparametric magnetic particle spectroscopy of
  {CoFe}$_{\textrm{2}}${O}$_{\textrm{4}}$ nanoparticles in viscous media.
\newblock {\em Journal of Physical Chemistry C}, 123(11):6787--6801, 2019.

\bibitem{enpuku_harmonic_2023}
Keiji Enpuku, Yi~Sun, Haochen Zhang, and Takashi Yoshida.
\newblock Harmonic signals of magnetic nanoparticle in suspension for
  biosensing application: Comparison of properties between {B}rownian- and
  {N}{\'e}el-dominant regions.
\newblock {\em Journal of Magnetism and Magnetic Materials}, 579:170878, 2023.

\bibitem{zverev_computer_2021}
Vladimir Zverev, Alla Dobroserdova, Andrey Kuznetsov, Alexey Ivanov, and
  Ekaterina Elfimova.
\newblock Computer simulations of dynamic response of ferrofluids on an
  alternating magnetic field with high amplitude.
\newblock {\em Mathematics}, 9(20):2581, 2021.

\bibitem{MRS74}
M.~A. Martsenyuk, Yu.~L. Raikher, and M.~I. Shliomis.
\newblock On the kinetics of magnetization of suspension of ferromagnetic
  particles.
\newblock {\em Zh. Eksp. Teor. Fiz.}, 65:834, 1973.
\newblock [Sov. Phys. JETP 38, 413 (1974)].

\bibitem{barrera_magnetization_2022}
Gabriele Barrera, Paolo Allia, and Paola Tiberto.
\newblock Magnetization dynamics of superparamagnetic nanoparticles for
  magnetic particle spectroscopy and imaging.
\newblock {\em Physical Review Applied}, 18(2):024077, 2022.

\bibitem{biederer_magnetization_2009}
S~Biederer, T~Knopp, T~F Sattel, K~L{\"u}dtke-Buzug, B~Gleich, J~Weizenecker,
  J~Borgert, and T~M Buzug.
\newblock Magnetization response spectroscopy of superparamagnetic
  nanoparticles for magnetic particle imaging.
\newblock {\em Journal of Physics D: Applied Physics}, 42(20):205007, 2009.

\bibitem{Shlio01}
M.~I. Shliomis.
\newblock Ferrohydrodynamics: Testing a third magnetization equation.
\newblock {\em Phys. Rev. E}, 64:060501, 2001.

\bibitem{EmbsLuecke_rigidrot}
J.~P. Embs, S.~May, C.~Wagner, A.~V. Kityk, A.~Leschhorn, and M.~L{\"u}cke.
\newblock Measuring the transverse magnetization of rotating ferrofluids.
\newblock {\em Phys. Rev. E}, 73:036302, 2006.

\bibitem{weng_magnetoviscosity_2013}
Huei~Chu Weng, Cha'o-Kuang Chen, and Min-Hsing Chang.
\newblock Magnetoviscosity in magnetic fluids: Testing different models of the
  magnetization equation.
\newblock {\em Smart Science}, 1(1):51--58, 2013.

\bibitem{gradsteyn}
I~Gradshteyn and I~Ryzhik.
\newblock {\em Table of Integrals, Series and Products}.
\newblock Academic Press, London, 7th edition, 2007.

\bibitem{IKH01}
P.~Ilg, M.~Kr{\"o}ger, and S.~Hess.
\newblock Magnetoviscosity and orientational order parameters of dilute
  ferrofluids.
\newblock {\em J. Chem. Phys.}, 116(20):9078--9088, 2002.

\bibitem{Deissler2014}
Robert~J. Deissler, Yong Wu, and Michael~A. Martens.
\newblock Dependence of {Brownian} and {N}\'eel relaxation times on magnetic
  field strength.
\newblock {\em Med. Phys.}, 41(1):012301, 2014.

\bibitem{Coffey:1993}
W~T Coffey, P~J Cregg, and Yu~P Kalmykov.
\newblock {On the theory of Debye and N{\'e}el relaxation of single domain
  ferromagnetic particles}.
\newblock In I~Prigogine and S~A Rice, editors, {\em Advances in Chemical
  Physics}, volume LXXXIII, pages 263--464. John Wiley \& Sons, New York, 1993.

\bibitem{ilg:DJ2023}
Patrick Ilg.
\newblock {Nonequilibrium response of magnetic nanoparticles to time--varying
  magnetic fields: contributions from Brownian and N{\'e}el processes}.
\newblock {\em Phys. Rev. E}, 109:034603, March 2024.

\bibitem{egler-kemmerer_real-time_2022}
Alexander-N. Egler-Kemmerer, Abdulkader Baki, Norbert L{\"o}wa, Olaf Kosch,
  Raphael Thiermann, Frank Wiekhorst, and Regina Bleul.
\newblock Real-time analysis of magnetic nanoparticle clustering effects by
  inline-magnetic particle spectroscopy.
\newblock {\em Journal of Magnetism and Magnetic Materials}, 564:169984, 2022.

\bibitem{henry_ultra-precise_2023}
Manus Henry.
\newblock An ultra-precise {Fast Fourier Transform}.
\newblock {\em Measurement}, 220:113372, 2023.

\end{thebibliography}
\end{document}